\documentclass[aps,prd,showpacs,notitlepage,nofootinbib,floatfix,showkeys]{revtex4-2}

\usepackage{blindtext}
\usepackage{hyperref}
\usepackage{amsmath,amssymb}
\usepackage{float}
\usepackage{microtype}

\usepackage{graphicx}
\usepackage{bm}
\usepackage{latexsym}
\usepackage{epsfig}
\usepackage{psfrag}
\usepackage{color}
\usepackage[dvipsnames]{xcolor}
\usepackage{subfigure}
\usepackage{graphicx}

\usepackage{amssymb}
\usepackage{amsmath}
\usepackage{bm}
\usepackage{latexsym}
\usepackage{epsfig}
\usepackage{psfrag}
\usepackage{amsmath}
\usepackage[normalem]{ulem}
\usepackage{textcomp}
\usepackage{color}
\usepackage{pstricks}
\usepackage[utf8]{inputenc}
\usepackage{array}

\def\nn{\nonumber} 

\def\f{\frac}
\def\l{\left}
\def\r{\right}
\def\d{{\rm d}}
\def\Mpl{M_{_{\mathrm{Pl}}}}
\def\mpcinv{{\rm Mpc}^{-1}}

\def\ps{\mathcal{P}_{_{\mathrm{S}}}}
\def\pc{\mathcal{P}_{_{\mathrm{C}}}}

\def\ns{n_{_{\mathrm{S}}}}

\def\cR{{\mathcal R}}

\def\cRG{{\mathcal R}^{^{_{\rm G}}}}
\def\cRk{{\mathcal R}_{\bm k}}
\def\cRGk{{\mathcal R}^{^{_{\rm G}}}_{\bm k}}

\def\fnl{f_{_{\rm NL}}}

\def\vka{{\bm k}_{1}}
\def\vkb{{\bm k}_{2}}
\def\vkc{{\bm k}_{3}}

\def\vk{{\bm{k}}}

\def\d{{\mathrm{d}}}

\allowdisplaybreaks

\begin{document}
\title{Indirect imprints of primordial non-Gaussianity on cosmic microwave background}
\author{Barnali Das}
\email{bd18ms201@iiserkol.ac.in} 
\affiliation{Department of Physical Sciences, 
Indian Institute of Science Education and Research Kolkata, 
Mohanpur, Nadia~741246, India}
\author{H.~V.~Ragavendra}
\email{ragavendra@rrimail.rri.res.in}
\affiliation{Raman Research Institute, C.~V.~Raman Avenue, Sadashivanagar, 
Bengaluru~560080, India}

\begin{abstract}
Primordial non-Gaussianity arising from inflationary models is a unique probe
of non-trivial dynamics of the inflaton field and its interactions with other
fields. Often when examining and constraining the scalar non-Gaussianity arising
from inflation, certain templates are adopted for the scalar non-Gaussianity 
parameter $\fnl$, in classifying their behaviors in terms of wavenumbers. The 
current constraints from cosmic microwave background (CMB) on such templates of 
$\fnl$ are weak and provide rather large bounds on their amplitudes. In this work, 
we explore a different method of constraining $\fnl$ through their effect on the 
scalar power. We compute the correction to the scalar power due to $\fnl$ while 
accounting for its generic scale dependence. 
We then compute the angular power spectrum of CMB arising from such non-Gaussian 
corrections to explore possible imprints. 
We initially illustrate this method using the conventional templates of $\fnl$ 
such as local, equilateral and orthogonal types, with and without the running of 
the parameter. We further employ this method to an oscillatory form of $\fnl$ and 
lastly on a realistic model of inflation proposed by Starobinsky. Though this 
method does not improve much on the constraints on the first three templates of 
$\fnl$, it provides interesting insights on models that do not conform to these 
templates. We infer that the non-Gaussian correction to the spectrum can be 
sensitive to model parameters that are degenerate at the level of the original 
power spectrum. Hence, this method of computing indirect imprints of $\fnl$ on 
angular power spectrum of CMB provides a new avenue to explore primordial scalar 
non-Gaussianity and possibly constrain them effectively.
\end{abstract}

\maketitle


\section{Introduction}

Models of inflation that give rise to a variety of non-trivial features in the 
primordial scalar power spectrum have been studied widely in the 
literature~(for some of the earlier efforts, 
refer~\cite{Contaldi:2003zv,Sinha:2005mn,Hazra:2013nca,Meerburg:2013dla}, 
and for recent efforts, see~\cite{Ragavendra:2020old,Sohn:2022jsm,Braglia:2022ftm}). 
They are sought for their promise of improvement in the 
fit to the data of anisotropies in the cosmic microwave background at the level 
of angular power spectrum~\cite{Ade:2015lrj,Planck:2018jri}. 
Such features arise primarily due to non-trivial dynamics of the evolution of
the inflaton field, in the context of canonical single field inflationary models.
These models also give rise to scalar non-Gaussianities of large amplitudes and
non-trivial shapes and they have been explored at the level of 
bispectrum~\cite{Chen:2008wn,Martin:2011sn,Adshead:2011jq,Hazra:2012yn,Basu:2019jew,
Ragavendra:2019mek,Clarke:2020znk,Ragavendra:2023ret}.

While features at the level of power spectrum have been well constrained against
CMB dataset, the constraints are relatively weaker at the level of bispectrum.
The scalar non-Gaussianity parameter associated with the bispectrum, $\fnl$,
is typically parametrized using certain well-motivated templates, called local, 
equilateral and orthogonal templates, which are supposed to represent broad classes
of models~\cite{Babich:2004gb,Bartolo:2004if,Creminelli:2005hu,Komatsu:2010hc}. 
These templates provide ease of comparison against data and the current constraints
on the amplitudes of $\fnl$ of the respective templates are 
$\fnl^{\rm loc}=-0.9\pm 5.1$, $\fnl^{\rm eq} = -26 \pm 47$ and 
$\fnl^{\rm ortho} = -38 \pm 24$ at $1-\sigma$ level~\cite{Planck:2019kim}.
These large bounds, obtained with the current level of sensitivity of the Planck
mission, certainly rule out exotic models of inflation leading to large
non-Gaussianity but do not inform much about the typical models and the associated
parameters that govern the scalar bispectrum.
Moreover, we should note that these simple templates are not sufficient to 
describe the non-trivial shapes of $\fnl$ that arise in models that lead to 
features in the power spectrum. Hence, it is challenging to compare the
exact shape and behavior of $\fnl$ arising out of realistic models of inflation
directly against data and obtain constraints on the model parameters determining
the scalar non-Gaussianity.

In the absence of data and analysis techniques with better constraining power, 
we explore a method to examine $\fnl$ through its indirect effect on the angular 
power spectrum of CMB. We calculate the non-Gaussian correction to the primordial 
scalar power spectrum due to $\fnl$~\cite{Ragavendra:2020sop,Ragavendra:2020vud}.
Further, we compute the corresponding CMB angular spectrum and compare it against
the original spectrum due to Gaussian perturbations.
An important advantage of this method is that, it accounts for the complete and
arbitrary scale dependence of $\fnl$ without relying on templates.
Also, it is computationally simpler than a direct of comparison of $\fnl$ against
data. Therefore, it is robust enough to be employed for any given model of 
inflation and arrive at possibly better bounds on $\fnl$ and the relevant model 
parameters. 

This approach is equivalent to computing the one-loop correction to the spectrum 
when the modes of perturbations are unaffected in the super-Hubble regime, which is 
especially true for the large scales of CMB. 
There have been several efforts in the context of computing loop-corrections to 
primordial spectrum~\cite{Sloth:2006az,Byrnes:2007tm,Cogollo:2008bi,
Dimastrogiovanni:2008af,Senatore:2009cf}.
We should also note that there is considerable interest in recent literature in
computing such loop-level non-Gaussian contributions to the scalar power, especially 
in the context of enhancing scalar power over small scales in ultra slow roll models 
of inflation~\cite{Ragavendra:2020sop,Kristiano:2021urj,Kristiano:2022maq,
Riotto:2023hoz,Kristiano:2023scm,Choudhury:2023rks,Firouzjahi:2023aum}.
Such loop corrections have interesting implications for the case of primordial
perturbations evolving from excited initial states~\cite{Ragavendra:2020vud,Ota:2022xni}.
There are also efforts to study loop-level contribution to the spectrum of $21$ cm 
signal~\cite{Yamauchi:2022fri}, and to the spectral density of scalar-induced 
secondary gravitational waves~\cite{Cai:2018dig,Unal:2018yaa,
Cai:2019amo,Adshead:2021hnm,Ragavendra:2021qdu,Chen:2022dah}.
In this work, we utilize this method to compute possible imprints of $\fnl$ on 
the angular spectrum of CMB and study the possibility of effectively constraining 
the relevant model parameters.

The article is organized as follows. In Sec.~\ref{sec:pc}, we outline the 
computational scheme to calculate the correction to the scalar power
due to $\fnl$.
In Sec.~\ref{sec:templates}, we illustrate our method using the typical templates 
of $\fnl$, with and without a running that may be present in these templates.
We also employ it on an oscillatory template of $\fnl$ which gives us
more insight into spectra with non-trivial scale dependences.
In Sec.~\ref{sec:staro}, we employ the method to a realistic model of inflation 
that was originally proposed by Starobinsky. Lastly, we conclude with a summary
and outlook in Sec.~\ref{sec:conc}.

As to the notations used in the article, we denote the original scalar power
spectrum as $\ps(k)$ and the non-Gaussian correction to the spectrum as $\pc(k)$.
We use $\fnl(k_1,k_2,k_3)$ to denote the complete form of the scalar non-Gaussianity 
parameter, while denote the amplitudes of parameter in specific templates as
$\fnl^{\rm type}$, where `${\rm type}$' may be ${\rm loc}$, eq, ${\rm ortho}$
or ${\rm osc}$, denoting the respective templates.


\section{Correction to the power spectrum due to $\fnl(k_1,k_2,k_3)$}
\label{sec:pc}
In this section, we shall setup the method of calculating the correction 
to the primordial scalar power spectrum arising due to $\fnl(k_1,k_2,k_3)$. 
We shall briefly outline the detailed derivation as presented in 
Ref.~\cite{Ragavendra:2021qdu} and also closely follow the conventions therein.
We consider the mode function associated with the primordial curvature 
perturbation, namely $\cRk$ and expand it in Fourier space to include the 
related non-Gaussianity as
\begin{eqnarray}
\cRk(\eta) &=& \cRGk(\eta) - \f{3}{5}\int\f{\d^3\vka}{(2\,\pi)^{3/2}}
\cRG_{\bm{k_1}}(\eta)\cRG_{\bm{k}-\bm{k_1}}(\eta)\,
\fnl[\bm{k},(\vka-\bm{k}),-\vka]\,,
\label{eq:fnl-gen}
\end{eqnarray}
where $\eta$ denotes the conformal time and $\cRGk$ denotes the Gaussian part of $\cRk$. 
Such an expansion has been used in different contexts in 
literature~\cite{Schmidt:2010gw,Agullo:2021oqk}.
This expansion generalizes the conventional expansion of curvature perturbation in
real space as $\cR(\bm{x},\eta) = \cRG(\bm{x},\eta) - (3/5)\fnl[\cRG(\bm{x},\eta)]^2\,,$ 
where $\fnl$ is assumed to be local without any scale-dependence (see, for instance, 
Refs.~\cite{Maldacena:2002vr,Martin:2011sn}).
We are interested in computing $\cRk(\eta)$ at $\eta_{\rm e}$ close to the end of
inflation. There can be additional time dependence in this expansion through$\fnl$. 
But for the large scales of our interest, the modes at $\eta_{\rm e}$ are in their 
super-Hubble regime. Even if there are brief departures from slow-roll during inflation, 
as long as there is a slow-roll phase prior to the end of inflation, these modes are 
largely unaffected and are essentially constant in their amplitudes. Therefore, the time 
dependence in $\fnl$ due to deviations from slow-roll between Hubble-exit of modes and 
$\eta_e$ shall be negligible.

To illustrate that $\fnl(k_1,k_2,k_3)$ captures the non-Gaussianity arising from the
scalar bispectrum, we can compute the two-point and three-point correlations of 
$\cRk(\eta)$ as expanded in Eq.~\eqref{eq:fnl-gen}. Using the definitions of scalar
power and bi-spectra, $\ps(k)$ and ${\cal B}(k_1,k_2,k_3)$~\cite{Maldacena:2002vr}
\begin{eqnarray}
\langle \cRG_{\vka}\cRG_{\vkb} \rangle &=& \f{2\pi^2}{k_1^3}\ps(k_1)
\delta^{(3)}(\vka+\vkb)\,, \\
\langle \cRG_{\vka}\cRG_{\vkb}\cRG_{\vkc} \rangle &=& (2\pi)^{3}\,{\cal B}(k_1,k_2,k_3)
\delta^{(3)}(\vka+\vkb+\vkc)\,,
\end{eqnarray}
we can show that the non-Gaussianity parameter $\fnl(k_1,k_2,k_3)$ can be expressed 
in terms of them as~\cite{Martin:2011sn,Hazra:2012yn,Ragavendra:2020old,Ragavendra:2021qdu,Ragavendra:2023ret}
\begin{eqnarray}
\fnl(k_1,k_2,k_3) &=& -\f{10\sqrt{(2\pi)}}{3}
\f{(k_1k_2k_3)^3{\cal B}(k_1,k_2,k_3)}
{\bigg[k_1^3\ps(k_2)\ps(k_3) + k_2^3\ps(k_1)\ps(k_3) + k_3^3\ps(k_1)\ps(k_2)\bigg]}\,.
\label{eq:fnl-ps-g}
\end{eqnarray}
The expectation values in the definitions used above are evaluated in the 
perturbative vacuum typically at $\eta_e$.
As we see, $\fnl(k_1,k_2,k_3)$ is a dimensionless ratio of bispectrum to
combination of power spectra. It is proportional to the scalar bispectrum 
${\cal B}(k_1,k_2,k_3)$ that receives its dominant contributions from the cubic order 
action governing $\cRk(\eta)$. 
Therefore $\fnl(k_1,k_2,k_3)$ captures the non-Gaussianity due to three-point
auto-correlation of $\cRk(\eta)$.

To compute the correction to the power spectrum $\pc(k)$ due to $\fnl(k_1,k_2,k_3)$, we 
compute the two point correlation of $\hat\cR_k(\eta)$ of Eq.~\eqref{eq:fnl-gen} as
\begin{eqnarray}
\langle \hat{\cR}_{\bm{k_1}} \hat{\cR}_{\bm{k_2}} \rangle &=& 
\langle \hat{\cal R}^{^{_{\rm G}}}_{\bm{k_1}} 
\hat{\cal R}^{^{_{\rm G}}}_{\bm{k_2}} \rangle
+ \f{9}{25}\int\f{\d^3\vka'}{(2\,\pi)^{3}}\int \d^3\vkb'
\langle \hat{\cal R}^{^{_{\rm G}}}_{\bm{k'_1}} 
\hat{\cal R}^{^{_{\rm G}}}_{\bm{k_1}-\bm{k'_1}} 
\hat{\cal R}^{^{_{\rm G}}}_{\bm{k'_2}} 
\hat{\cal R}^{^{_{\rm G}}}_{\bm{k_2}-\bm{k'_2}} \rangle \nn \\
& & \times \,
\fnl\big(k_1 ,\vert \vka'-\vka \vert ,k_1' \big)\,
\fnl\big(k_2,\vert \vkb'-\vkb \vert , k_2' \big)\,.
\label{eq:Rk4}
\end{eqnarray}
Using Wick's theorem to express the four point correlation in terms of the two 
point correlations, the above equation becomes
\begin{eqnarray}
\ps^{\rm M}(k) &=& \ps(k) + 
\f{9}{50\pi} k^3 \int \d^3\vka\,\f{\ps(k_1)}{k_1^3}
\f{\ps(\vert\bm{k}-\vka\vert)}{\vert \bm{k}-\vka \vert^3}
\,\fnl^2[k,\vert \vka-\bm{k} \vert, k_1]\,,
\end{eqnarray}
where $\ps(k)$ denotes the original power spectrum corresponding to the Gaussian 
perturbations $\cRGk(\eta)$, $\ps^{\rm M}(k)$ is the complete modified spectrum and 
the correction $\pc(k)$ can be identified as
\begin{eqnarray}
\pc(k) &=& \f{9}{50\pi} k^3 \int \d^3\vka\,\f{\ps(k_1)}{k_1^3}
\f{\ps(\vert\bm{k}-\vka\vert)}{\vert \bm{k}-\vka \vert^3}
\,\fnl^2[{k},\vert \vka-\bm{k} \vert,k_1]\,.
\end{eqnarray}
Using suitable change of variables, we can rewrite the above integral to a
computationally convenient form of
\begin{eqnarray}
\pc(k) &=& \f{9}{25} \int_0^\infty \d x \int_{\vert 1-x \vert}^{\vert 1 + x \vert}
\d y \f{\ps(kx)}{x^2}\,\f{\ps(ky)}{y^2}\,\fnl^2(k,kx,ky)\,.
\label{eq:pc-fnl}
\end{eqnarray}
This is the correction to the scalar power spectrum that we shall compute for
various cases of $\fnl(k_1,k_2,k_3)$. 

At this stage, a few points have to be noted regarding this integral.
The integral appears to have divergences at the points $(x,y)=(1,0)$ and
$(x,y)=(0,1)$. This divergence is essentially the point where the momentum
of integration $k_1 \to 0$ or $\bm{k}_1 \to \vk$. 
These divergences may not be an issue as long as the behavior of $\ps(k_1)$ or
$\fnl(k,k_1,\vert \vk - \vka \vert)$ decay as $k_1$ or faster as $k_1 \to 0$.
In the absence of such a curtailing nature of $\ps(k_1)$ or
$\fnl(k,k_1,\vert \vk - \vka \vert)$, these divergences can be regulated by 
introduction of a $k_{\rm min}$. 
To choose a model-independent value of $k_{\rm min}$, we shall set 
$k_{\rm min}/\,\mpcinv=10^{-6}$ which corresponds to the scale when the
largest scale observable today was sufficiently deep inside the Hubble radius.
Besides, due to the symmetric nature of the integrand under the exchange of 
$\vka$ and $\vk-\vka$, or equivalently $x$ and $y$, regulating one divergence 
automatically takes care of the other one.
For discussions regarding this divergence, see Refs.~\cite{Sloth:2006az,Seery:2010kh,
Gerstenlauer:2011ti}.
Nevertheless, as we shall see, except for certain cases, the nature of $\fnl$
provides us the necessary regulation of the integrands so that there arise
no divergence in the computation of these integrals.


\section{Templates of interest}\label{sec:templates}

In this section, we shall consider various templates of $\fnl(k_1,k_2,k_3)$ that
are typically considered in inflationary literature, namely local, orthogonal 
and equilateral templates. We shall introduce running of $\fnl(k_1,k_2,k_3)$ 
through the parameter $n_{\rm NG}$. 
Though the method outlined works for arbitrary $\fnl(k_1,k_2,k_3)$, we shall 
initially illustrate the usage with these templates. Later, we shall use this 
method on an oscillatory $\fnl(k_1,k_2,k_3)$.
We shall use the publicly available package called~{\tt CAMB} to compute the moments
of the angular spectrum of CMB arising out of $\pc(k)$ for each case along with 
the respective $\ps(k)$~\cite{Lewis:1999bs}.
Since, we wish to examine the broad features and amplitude of the angular spectra 
due to $\pc(k)$ against the spectra due to $\ps(k)$,
we have turned off non-linear lensing in our calculation of $C_\ell$s which causes
minor difference mainly over large values of $\ell$. 


\subsection{Local type}\label{subsec:local}
The local type of $\fnl(k_1,k_2,k_3)$ is a widely studied template and has been
known to characterize scalar non-Gaussianity arising out of multi-field models
and in ultra slow roll models around the peak amplitude of scalar 
power~\cite{Jung:2016kfd,Planck:2019kim,Atal:2018neu,Ragavendra:2023ret}.
In this case, $\fnl(k_1,k_2,k_3)$ is taken to be a constant without any scale
dependence. We shall refer to it as $\fnl^{\rm loc}$. We shall take the original
scalar power spectrum to be the nearly scale-invariant form of 
\begin{eqnarray}
\ps(k) &=& A_s(k)\l(\f{k}{k_\ast}\r)^{\ns-1}.
\label{eq:ps-ns}
\end{eqnarray}
Using Eq.~\eqref{eq:pc-fnl}, we obtain the corresponding $\pc(k)$ to be
\begin{eqnarray}
\pc^{\rm loc}(k) &=& -\f{18}{25}\,\ps^2(k)\,(\fnl^{\rm loc})^2\,\left(1 + 2\lim_{k_{\rm min} \rightarrow 0} \ln{\f{k_{\rm min}}{k}} \right).
\end{eqnarray}
For details of computation of the integral see App.~\ref{app:integrands}.
Note that, the second term in parentheses is dominant and evaluates to a negative 
value. So, the overall value of $\pc(k)$ always remains positive.
Besides, $\pc^{\rm loc}(k)$ diverges as $k_{\rm min} \rightarrow 0$, so we take 
a non-zero $k_{\rm min}$ for computing the integral as mentioned earlier. 
The range of wavenumbers corresponding to the CMB scales probed by Planck is 
$k/\,\mpcinv = 10^{-4}$ to $k/\,\mpcinv = 1$~\cite{Planck:2018jri}. We assume the 
lowest wavenumber corresponding to the largest scale to evolve when $k = 100aH$, 
where $aH$ is the inverse of comoving Hubble radius. Hence, we choose the value of 
$k_{\rm min}/\,\mpcinv = 10^{-6}$. We shall use the same value for $k_{\rm min}$ 
for any subsequent case that contains a divergence in the integral.

\begin{figure}[t]
\centering
\includegraphics[width=8.75cm]{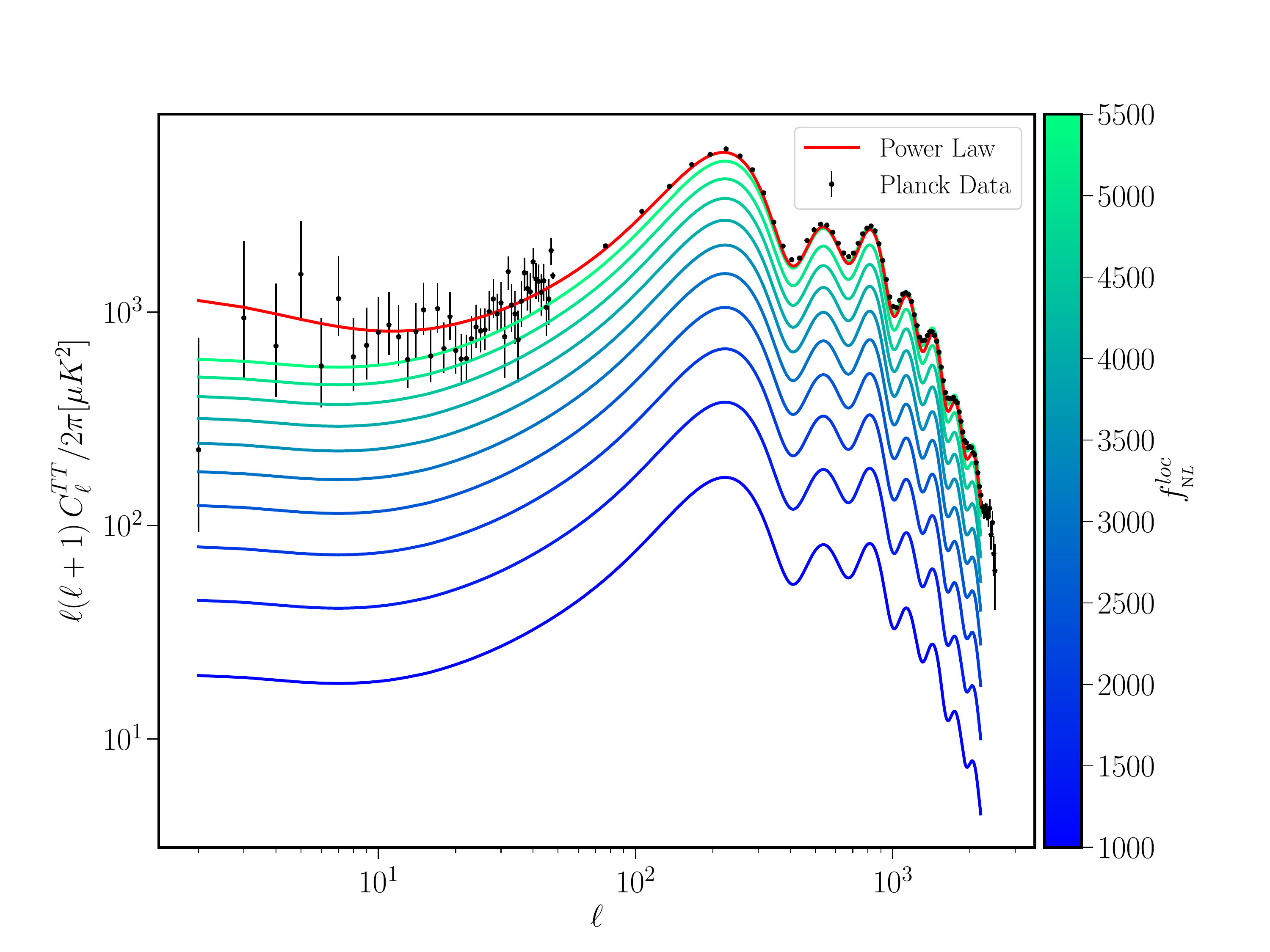}
\includegraphics[width=8.75cm]{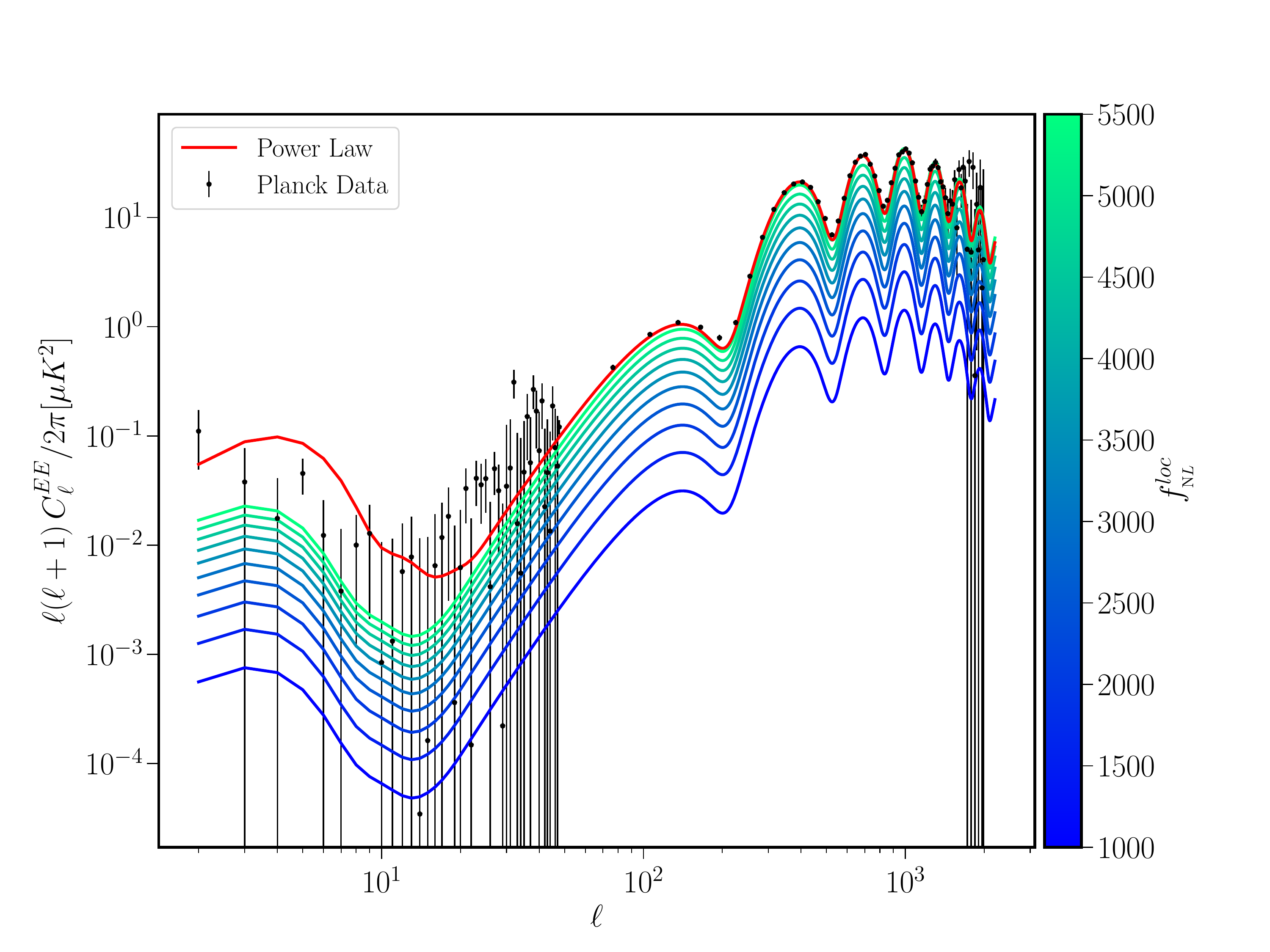}
\includegraphics[width=8.75cm]{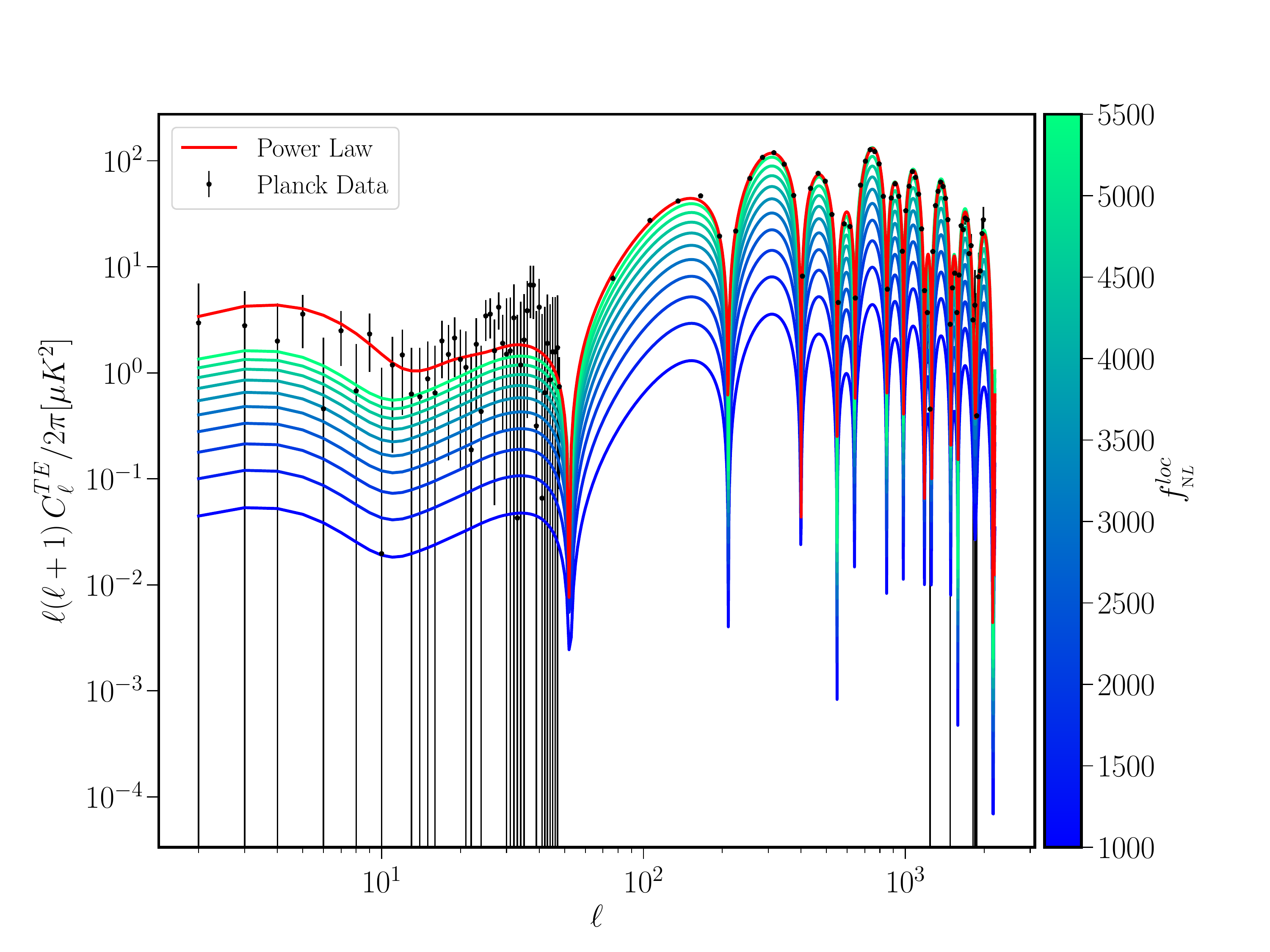}
\vskip -0.1in
\caption{We plot the angular spectra, $C_\ell$s, due to $\pc^{\rm loc}(k)$ across a range of the parameter $\fnl^{\rm loc}$. We illustrate all the three angular spectra arising due to scalar power namely $TT, TE$ and $EE$. The red curves represent the respective $C_\ell$s due to $\ps(k)$ which is the power law. We also plot the data points of Planck 2018 in black dots along with errorbars. The amplitude of $C_{\ell}$s due to $\pc^{\rm loc}(k)$, plotted in shades of blue to green as indicated in the colorbars, increase with increase in the value of $\fnl^{\rm loc}$. We find that the value of $\fnl^{\rm loc}$ required to get amplitudes comparable to the standard $C_{\ell}$s is $\sim 5000$.}
\label{fig:Cls_loc}
\end{figure}
Utilizing the $\ps(k)$ and $\pc^{\rm loc}(k)$, we compute the respective CMB 
angular spectra using {\tt CAMB}. Since we focus on the scalar power and the 
associated non-Gaussian correction we compute just the three related correlations 
of the CMB angular spectrum, namely $TT, TE$ and $EE$ correlations.
For subsequent templates and models, we shall focus only on $TT$ correlations
to examine the effects of corresponding $\pc(k)$.
In Fig.~\ref{fig:Cls_loc}, we present the angular spectra obtained from 
$\pc^{\rm loc}(k)$ along with those obtained from $\ps(k)$. Note that these 
are separate contributions from the original and non-Gaussian parts of the 
power spectrum and the complete power spectrum shall be the sum of both.
We find that the amplitude of $C_\ell$s increases as $\fnl^{\rm loc}$ is 
increased. But the value of $\fnl^{\rm loc}$ required to give rise to
$C_\ell$s of magnitude comparable to the original spectrum is around $5000$.
This value is evidently much higher than the bound obtained from direct 
constraint, i.e. $\fnl^{\rm loc} = -0.9 \pm 5.1$. Therefore, we find that
our method is not particularly useful in constraining this template.


\subsection{Orthogonal type}\label{subsec:ortho}
We turn to the template of orthogonal type $\fnl(k_1,k_2,k_3)$ which is known to
arise in models with non-Bunch Davies initial states for 
perturbations~\cite{Meerburg:2009ys,Brandenberger:2012aj,Ragavendra:2020vud}.
The scalar bispectrum for this case is parametrized as~\cite{Planck:2019kim}
\begin{eqnarray} 
{\cal B}^{\rm ortho}\left(k_1, k_2, k_3\right) &=& \f{6 A_s^2 \fnl^{\rm ortho}}{k_*^{2(\ns-1)}}\left\{-\f{3}{\left(k_1 k_2\right)^{4-\ns}}-\f{3}{\left(k_2 k_3\right)^{4-\ns}}-\f{3}{\left(k_3 k_1\right)^{4-\ns}} -\f{8}{\left(k_1 k_2 k_3\right)^{\f{2}{3}\left(4-\ns\right)}}\right.\nn \\
& & \left.+\left[\f{3}{k_1^{\left(4-\ns\right) / 3} k_2^{2\left(4-\ns\right) / 3} k_3^{4-\ns}}+5 \text { permutations }\right]\right\}.
\end{eqnarray}
The original scalar power is assumed to be the nearly scale invariant spectrum as
in Eq.~\eqref{eq:ps-ns}.
So, the form of the non-Gaussianity parameter [cf.~Eq.~\eqref{eq:fnl-ps-g}], in 
terms of the variables of integration $x,y$ as it appears in Eq.~\eqref{eq:pc-fnl},
becomes
\begin{eqnarray}
    \fnl(k, k x, k y) &=& \f{160\sqrt{2\pi}\fnl^{\rm ortho}x^3y^3}{1+x^3+y^3}\left\{\f{1}{x^2 y^2} + \f{3}{8}\left(\f{1}{x^3}+\f{1}{y^3}+\f{1}{x^3 y^3}\right) - \f{3}{8}\left(\f{1}{x^3 y}+\f{1}{x y^3}+\f{1}{x^2 y}+\f{1}{x y^2}+\f{1}{x^3 y^2}+\f{1}{x^2 y^3}\right)\right\}.\;\;\;\;\;\;
\end{eqnarray}
In the above expression, we have ignored the tilt in $x$ and $y$ in the above, i.e. 
we have set $x^{\ns-1} = y^{\ns-1}=1$.  
For our analysis to infer the dominant effect due to the crucial parameter 
$\fnl^{\rm ortho}$, these minor contributions do not change the results appreciably.
Evidently, the parameter $\fnl^{\rm ortho}$ that determines the strength of the 
non-Gaussianity, shall determine the amplitude of $\pc(k)$.
With nearly scale invariant $\ps(k)$ and $\fnl(k,kx,ky)$ above, we calculate the 
$\pc(k)$ to be
\begin{eqnarray}
    \pc^{\rm ortho}(k) &=& \f{9}{25}(160)^2(2\pi)(\fnl^{\rm ortho})^2 A_s^2 \left(\f{k}{k_*}\right)^{2(\ns - 1)}\int_0^{\infty} d x \int_{|1-x|}^{(1+x)} d y \f{1}{x^2 y^2}\f{x^6y^6}{(1+x^3+y^3)^2} \nn \\ 
    & & \times \left[\f{1}{x^2 y^2} + \f{3}{8}\left(\f{1}{x^3}+\f{1}{y^3}+\f{1}{x^3 y^3}\right) - \f{3}{8}\left(\f{1}{x^3 y}+\f{1}{x y^3}+\f{1}{x^2 y}+\f{1}{x y^2}+\f{1}{x^3 y^2}+\f{1}{x^2 y^3}\right)\right]^2.
\end{eqnarray}
We find that the integral does not diverge anywhere in the range of integration, 
unlike local type~(see App.~\ref{app:integrands} for the shape of the integrand). 
Instead it has a maximum value of ${1}/{25}$ at $x = y = 1/2$. The rise is sharp 
along $y = 1-x$, so we can approximate the integral involved in $\pc(k)$ as
\begin{eqnarray}
    \pc^{\rm ortho}(k) &=& \f{9}{25}(160)^2(2\pi)(\fnl^{\rm ortho})^2 \ps^2(k)\int_0^{\infty} d x \int_{|1-x|}^{(1+x)} d y \delta(y-(1-x)) \f{1}{x^2 y^2}\f{x^6y^6}{(1+x^3+y^3)^2} \nn \\ 
    & & \times \left[\f{1}{x^2 y^2} + \f{3}{8}\left(\f{1}{x^3}+\f{1}{y^3}+\f{1}{x^3 y^3}\right) - \f{3}{8}\left(\f{1}{x^3 y}+\f{1}{x y^3}+\f{1}{x^2 y}+\f{1}{x y^2}+\f{1}{x^3 y^2}+\f{1}{x^2 y^3}\right)\right]^2 \\
    &=& \f{9}{25}(160)^2(2\pi)(\fnl^{\rm ortho})^2 \ps^2(k)\int_0^{\infty} d x \f{1}{x^2 (1-x)^2}\f{x^6(1-x)^6}{(1+x^3+(1-x)^3)^2} \nn \\ 
    & & \times \left[\f{1}{x^2 (1-x)^2} + \f{3}{8}\left(\f{1}{x^3}+\f{1}{(1-x)^3}+\f{1}{x^3 (1-x)^3}\right) \right. \nn \\
    & & - \left. \f{3}{8}\left(\f{1}{x^3 (1-x)}+\f{1}{x (1-x)^3}+\f{1}{x^2 (1-x)}+\f{1}{x (1-x)^2}+\f{1}{x^3 (1-x)^2}+\f{1}{x^2 (1-x)^3}\right)\right]^2 \\
    &=& \f{9}{25}(160)^2(2\pi)(\fnl^{\rm ortho})^2 \ps^2(k)\int_0^{\infty} d x \f{1}{16(3x^2-3x+2)^2} \\
    &=& 1.84\times 10^4\pi\,(\fnl^{\rm ortho})^2\ps^2(k)\,{\cal I}^{\rm ortho}\,,
\end{eqnarray}
where ${\cal I}^{\rm ortho} \simeq 2.95\times 10^{-2}$\,. Hence,
\begin{eqnarray}
\pc^{\rm ortho}(k) &\simeq & 1.71\times 10^3\,\ps^2(k)\,(\fnl^{\rm ortho})^2.
\end{eqnarray}

\begin{figure}[t]
\centering
\includegraphics[width=12cm,height=8cm]{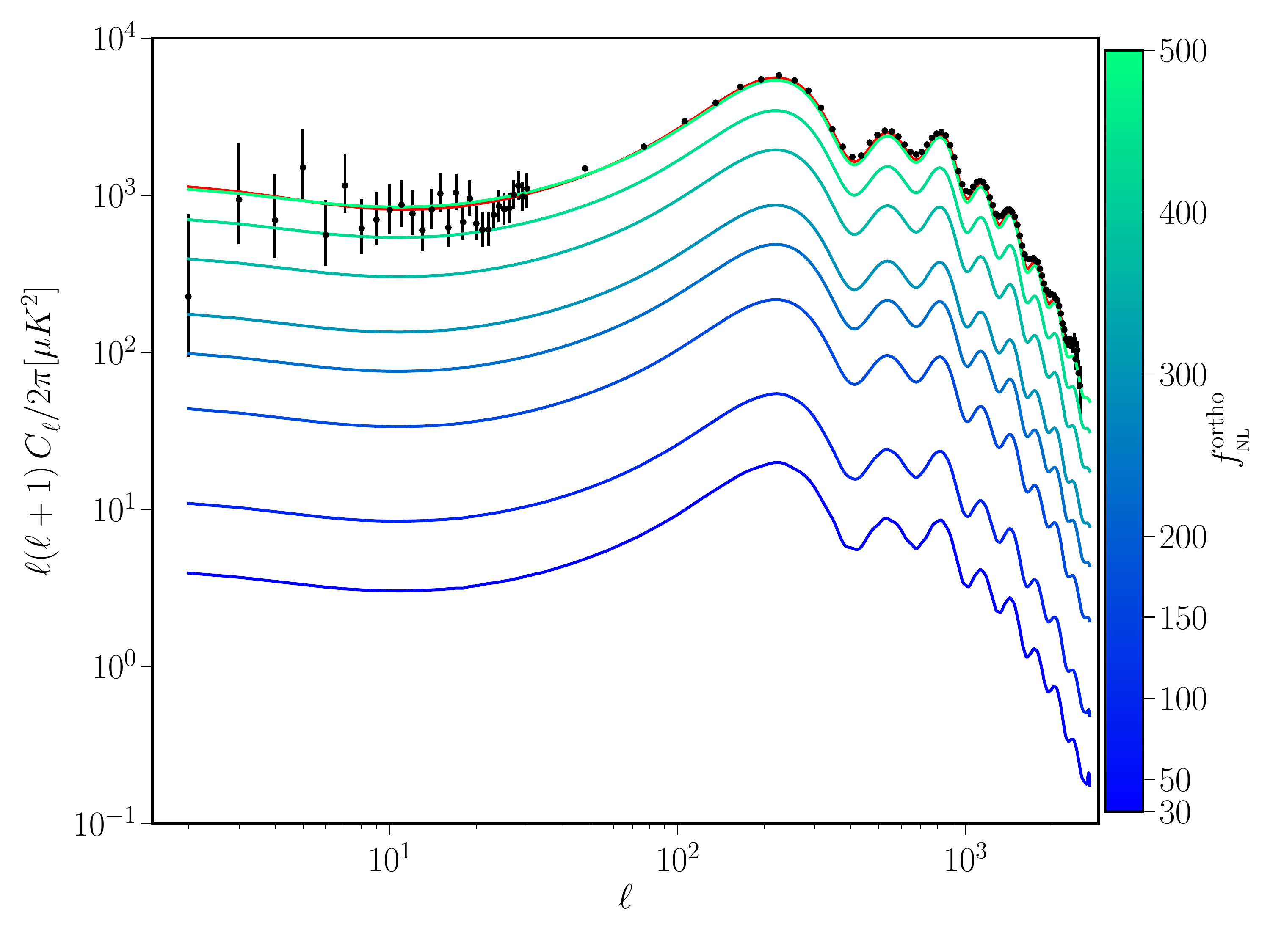}
\vskip -0.1in
\caption{The $TT$ angular power spectrum arising from $\pc^{\rm ortho}$ is 
presented (in shades of blue to green) across a range of the parameter 
$\fnl^{\rm ortho}$. The standard spectrum due to $\ps(k)$ is presented (in red)
along with the Planck data points (in black). As expected, the amplitude of 
$C_{\ell}$s due to $\pc^{\rm ortho}(k)$ decreases as $\fnl^{\rm ortho}$ is 
decreased. An interesting point to note is that the value of $\fnl^{\rm ortho}$ 
required to get amplitudes comparable to the standard $C_{\ell}$s is $\sim 500$, 
which is an order of magnitude lesser than it is required for $\fnl^{\rm loc}$.}
\label{fig:ClsTT_ortho_run_fnlortho}
\end{figure}
The CMB angular spectrum due to $\pc^{\rm ortho}(k)$ is presented in 
Fig.~\ref{fig:ClsTT_ortho_run_fnlortho}. We find that the amplitude of 
$\fnl^{\rm ortho}$ required to generate $C_\ell$s of amplitude comparable to 
the original spectrum is around $500$. This is an order of magnitude lesser 
than that of the local type. It is mainly due to the structure and numerical 
coefficients in the definition of the scalar bispectrum of this template. Yet, 
this value of $\fnl^{\rm ortho}$ is larger than the present constraint of 
$\fnl^{\rm ortho} = -39 \pm 64$ at $1$-$\sigma$ level. 


\subsection{Equilateral type}\label{subsec:equil}
We next consider $\fnl(k_1,k_2,k_3)$ of equilateral template which is representative
of non-Gaussianity arising from canonical single field inflationary models with
typical slow-roll evolution at early
times~\cite{Creminelli:2005hu,Hazra:2012yn,Ragavendra:2020old,Ragavendra:2020sop}. 
The scalar bispectrum for this case is parametrized as~\cite{Planck:2019kim}
\begin{eqnarray}
{\cal B}^{\rm eq}(k_1,k_2,k_3) &=& 6\,A_s^2\,\fnl^{\rm eq}\,
\bigg\{ -\frac{1}{(k_1k_2)^3}\,\left(\frac{k_1k_2}{k^2_\ast}\right)^{\ns-1}
-\frac{1}{(k_2k_3)^3}\,\left(\frac{k_2k_3}{k^2_\ast}\right)^{\ns-1}
-\frac{1}{(k_1k_3)^3}\,\left(\frac{k_1k_3}{k^2_\ast}\right)^{\ns-1}\nn \\
& & -\frac{2}{(k_1\,k_2\,k_3)^2}\,\left(\frac{k_1k_2k_3}{k^3_\ast}\right)^{2(\ns-1)/3}
+ \bigg[ \f{1}{k_1k_2^2k_3^3}\left(\f{k_1\,k_2^2k_3^3}{k_\ast^6}\right)^{(\ns-1)/3}
+~(5~{\rm permutations})\bigg]\bigg\}.
\end{eqnarray}
This template of bispectrum, along with the nearly scale invariant spectrum of 
Eq.~\eqref{eq:ps-ns}, when substituted in the expression of $\fnl$ in terms of 
$k, kx, ky$ as required in the integral of interest, gives us [cf.~Eq.~\eqref{eq:fnl-ps-g}]
\begin{eqnarray}
\fnl(k,k x,k y) &=& \f{20\sqrt{2\pi}\fnl^{\rm eq}x^3y^3}{(1+x^3+y^3)}
\bigg\{\f{1}{x^3} + \f{1}{y^3} + \f{1}{x^3y^3} + \f{2}{x^2y^2}
-\left[\f{1}{x^2y^3} +~(5~{\rm permutations})\right]\bigg\}\,.
\end{eqnarray}
This $\fnl(k,kx,ky)$ leads to $\pc(k)$ of the form
\begin{eqnarray}
\pc^{\rm eq}(k) = 288\pi\,\ps^2(k)\,(\fnl^{\rm eq})^2\,
\int_0^\infty {\rm d}x\int_{\vert 1-x \vert}^{1+x} {\rm d}y
& &\left(\f{1}{x^2y^2}\right)\,\f{x^6y^6}{(1+x^3+y^3)^2}
\bigg\{\f{1}{x^3} + \f{1}{y^3} + \f{1}{x^3y^3} + \f{2}{x^2y^2} \nn \\
& & -\left[\f{1}{xy^3} + \f{1}{x^3y} + \f{1}{xy^2} + \f{1}{x^2y}
+ \f{1}{x^2y^3} + \f{1}{x^3y^2}\right]\bigg\}^2\,,
\end{eqnarray}
where, as in the previous case, we have ignored the minor tilt arising due to 
$x^{\ns-1}$ and $y^{\ns-1}$ without losing much information about the amplitude 
or scale dependence of $\pc^{\rm eq}(k)$.
The above integrand does not diverge in the regime of integration and has a 
maximum of about $0.236$ at $(x,y)\simeq(0.719,0.719)$. Away from this maximum, 
the function quickly decays down~(see App.~\ref{app:integrands} for related details). 
So, we approximate the integral as this maximum value and integrate over a unit 
area around this point to obtain a simplified expression of $\pc^{\rm eq}(k)$
as
\begin{eqnarray}
\pc^{\rm eq}(k) &\simeq & 2.135 \times 10^{2}\,\ps^2(k)(\fnl^{\rm eq})^2\,.
\end{eqnarray}

\begin{figure}[t]
\centering
\includegraphics[width=12cm,height=8cm]{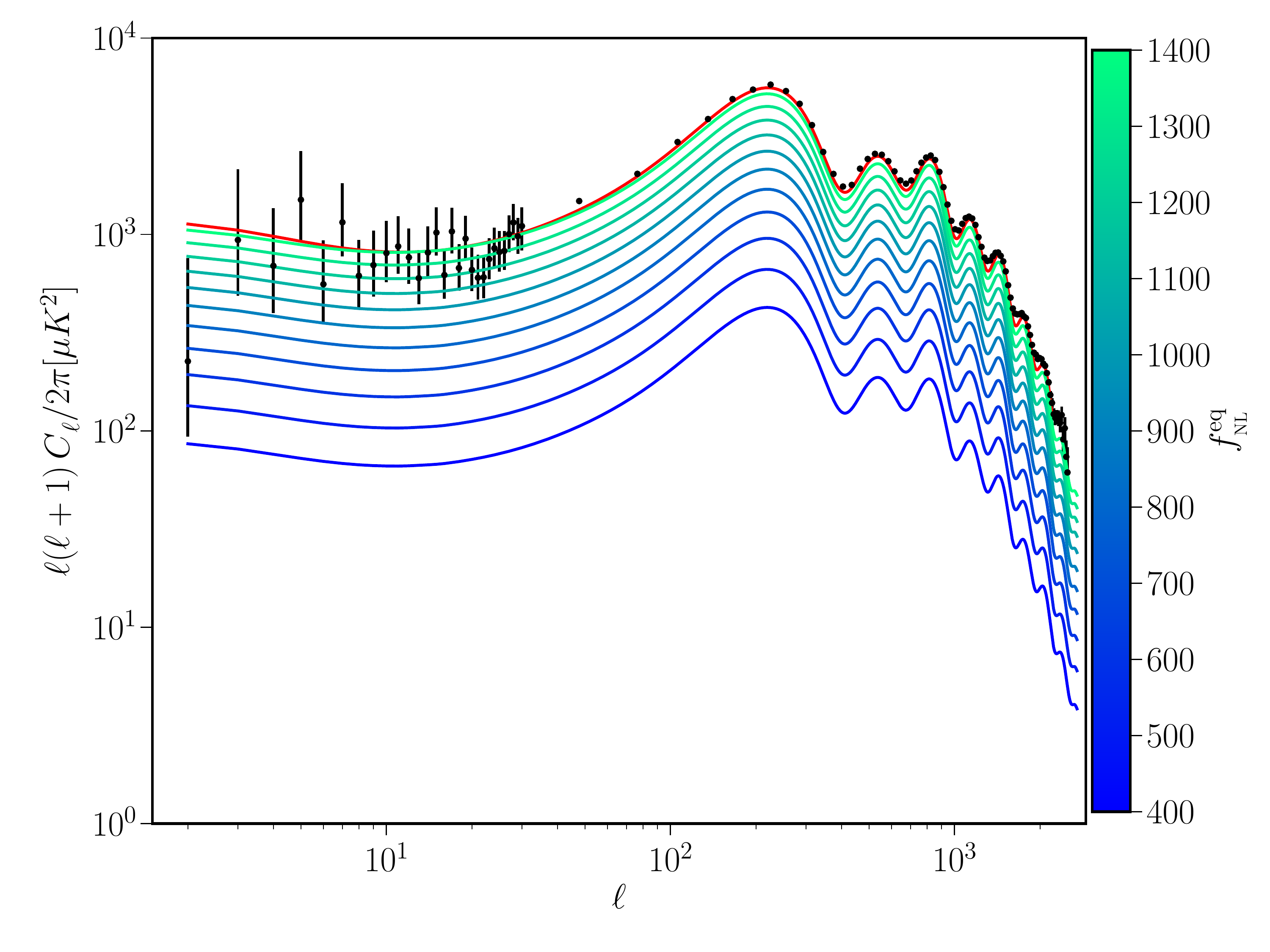}
\vskip -0.1in
\caption{The angular spectrum due to $\pc^{\rm eq}(k)$ is presented (in shades 
of blue to green) across a range of the parameter $\fnl^{\rm eq}$ along with
the standard spectrum due to $\ps(k)$ (in red) and the data points (in black).
At $\fnl^{\rm eq}\sim 1400$, the $C_\ell$ due to $\pc^{\rm eq}(k)$ become 
comparable to the amplitude of the standard $C_{\ell}$s. This is similar to the 
value that is required for $\fnl^{\rm loc}$.}
\label{fig:ClsTT_eq_run_fnleq}
\end{figure}
The CMB angular spectrum in this case is presented in 
Fig.~\ref{fig:ClsTT_eq_run_fnleq}. 
We find that the amplitude of $\fnl^{\rm eq}$ required to produce $C_\ell$s
of magnitude comparable to the standard $C_\ell$s due to $\ps(k)$ is around
$10^3$. Once again this value is higher than the direct constraint on the 
parameter, $\fnl^{\rm eq} = -26 \pm 47$ at $1$-$\sigma$ level. 
However, it is about a factor of $5$ lesser than the value of $\fnl^{\rm loc}$ 
required to lead to similar effect in $C_\ell$s.
Thus we see that the templates with weaker bounds through direct constraints
may be constrained better through our indirect method.


\subsection{Templates with running}\label{subsec:run}

Apart from the explicit scale dependence of the templates discussed above, a 
mild running of $\fnl(k_1,k_2,k_3)$ is typically be introduced as~\cite{Oppizzi:2017nfy,Planck:2019kim}
\begin{eqnarray}
\fnl^{\rm type-run}(k_1,k_2,k_3) &=& \fnl^{\rm type}(k_1,k_2,k_3)
\left( \frac{k_1 + k_2 + k_3}{3\,k_\ast}\right)^{n_{\rm NG}}\,,
\end{eqnarray}
where $n_{\rm NG}$ quantifies the running of the parameter about $k_\ast$ and
`type' can refer to one of local, equilateral or orthogonal templates. 
The typical value of $n_{\rm NG}$ is between $-0.1$ and $0.1$ and the constraints
on them are broad and consistent with zero~\cite{Planck:2019kim}.
In our method, it can be easily derived that introduction of such a running to
the templates modifies the corresponding $\pc(k)$ as
\begin{eqnarray}
\pc^{\rm type-run}(k) &=& \pc^{\rm type}(k)
\left(\f{k}{3\,k_\ast}\right)^{2n_{\rm NG}}\,.
\end{eqnarray}
Once again we have ignored the minor dependence of $n_{\rm NG}$ through the 
variables of integration, i.e. $x^{n_{\rm NG}},y^{n_{\rm NG}}$ in performing 
the integral and focus on capturing the prominent scale dependence through 
$k^{n_{\rm NG}}$. 

\begin{figure}[!ht]
\centering
\includegraphics[width=8cm]{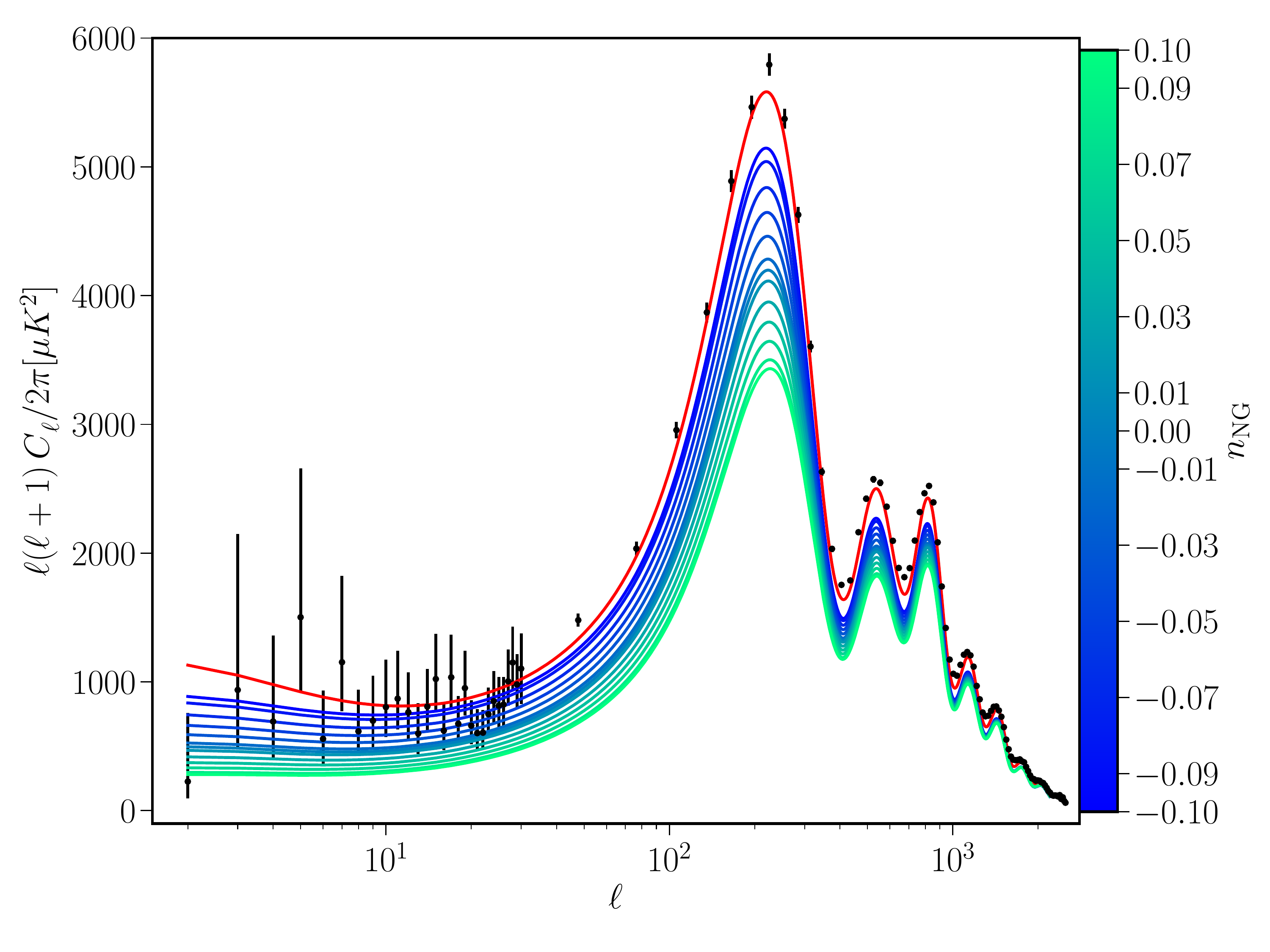}
\includegraphics[width=8cm]{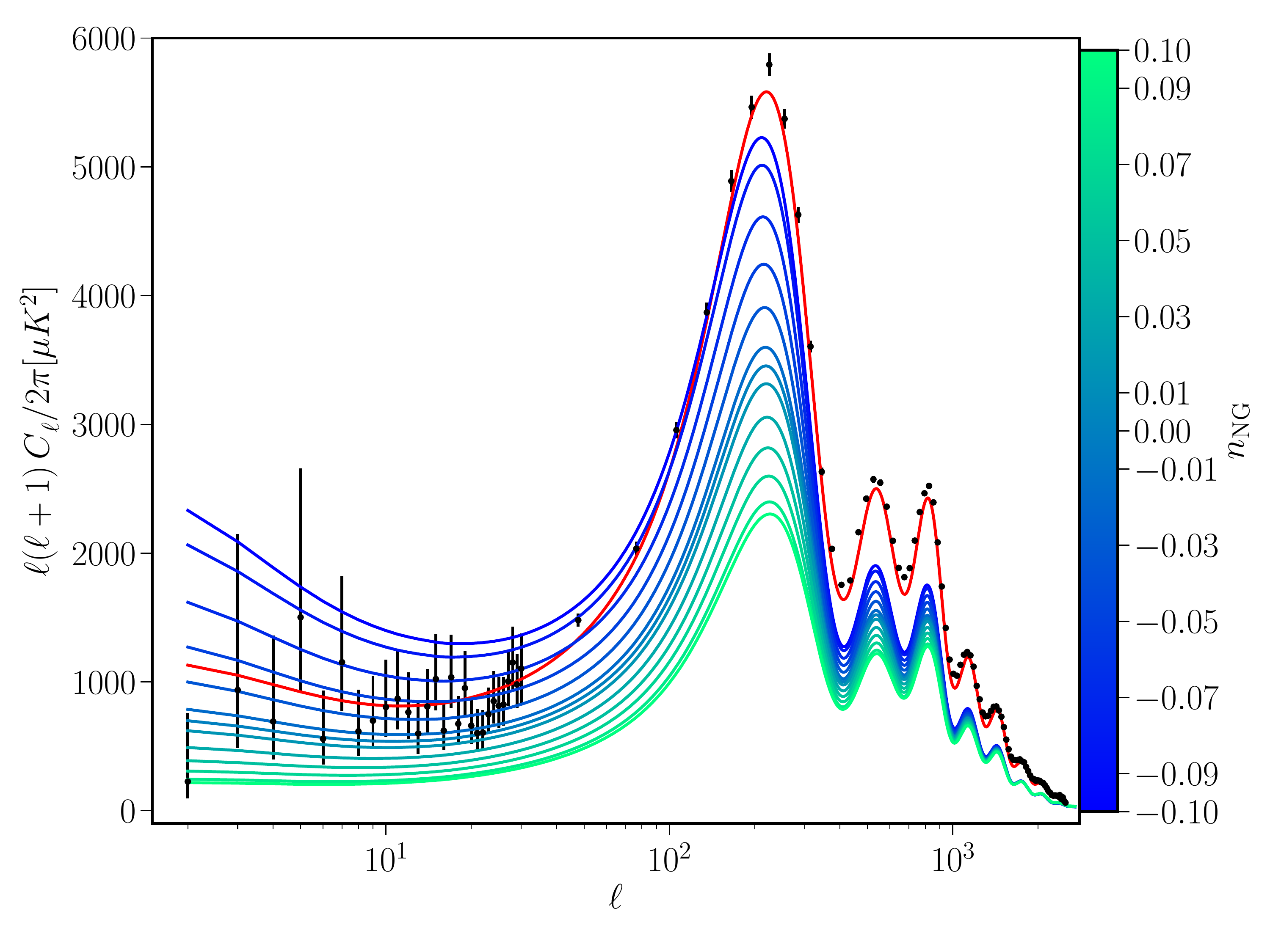}
\includegraphics[width=8cm]{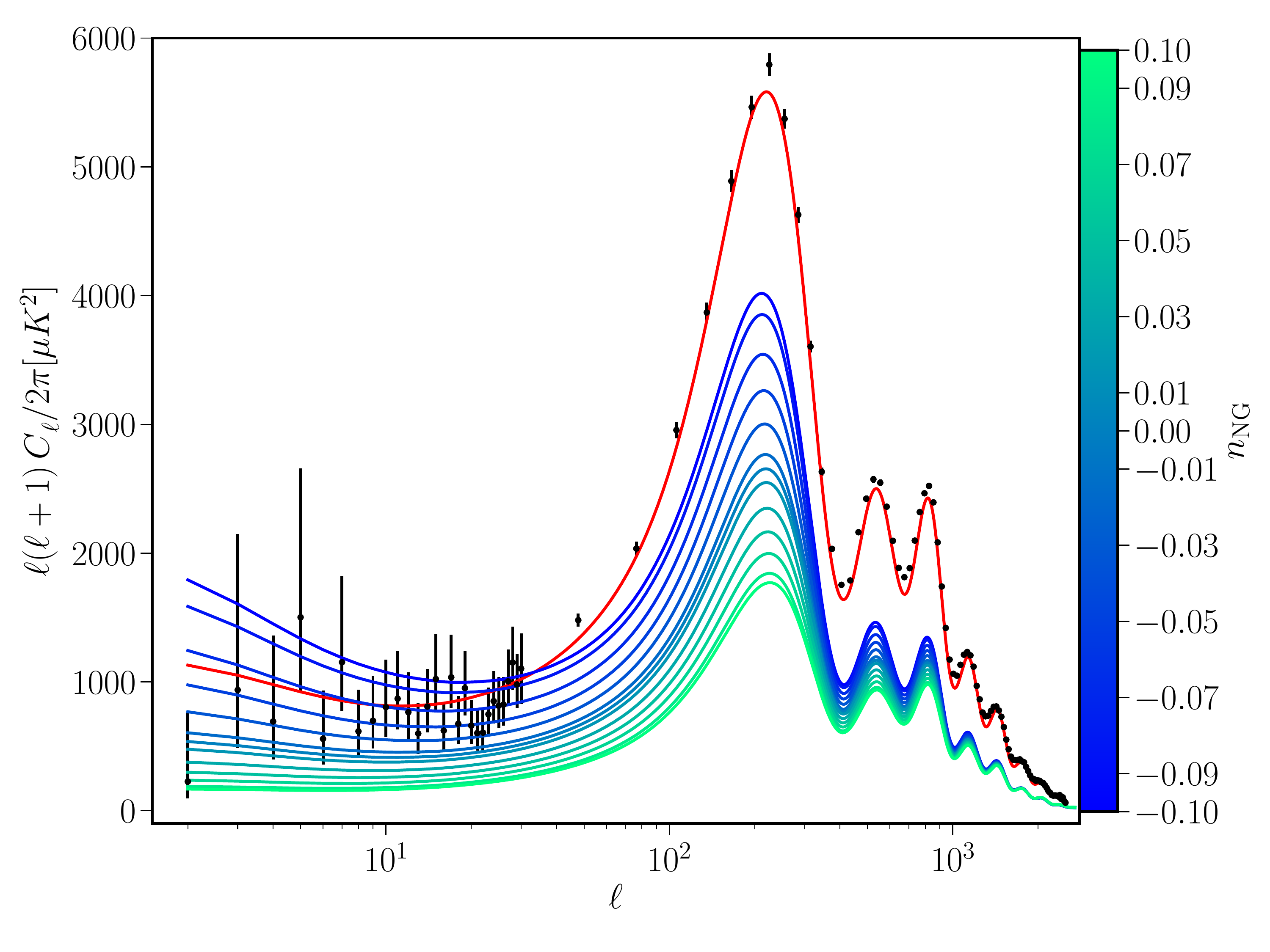}
\vskip -0.1in
\caption{The angular spectrum due to $\pc(k)$ is presented (in shades of blue to
green) for the case of running of $\fnl$ across a range of parameter $n_{\rm NG}$. 
We present it for the case of local (on top left), orthogonal (on top right) and 
equilateral (at the bottom) templates along with the standard $C_\ell$s due to 
power law $\ps(k)$ (in red) and data points (in black). 
We have fixed the respective $\fnl$ amplitudes to be $\fnl^{\rm loc}=5000$, 
$\fnl^{\rm ortho}=400$ and $\fnl^{\rm eq}=1000$ to illustrate the effect of 
$n_{\rm NG}$ on the $C_\ell$s.}
\label{fig:ClsTT_run_nng}
\end{figure}
The $C_\ell$s due to $\pc^{\rm type-run}(k)$ have been presented in 
Fig.~\ref{fig:ClsTT_run_nng}, for local, orthogonal and equilateral templates.
We fix the values of $\fnl^{\rm type}$ in respective templates and focus on the 
effect due to variation of ${n_{\rm NG}}$. We find that for a given template, 
${n_{\rm NG}}$ can alter the overall tilt of the angular spectrum and this in
turn affects the amplitude of $C_\ell$s over large scales.
Therefore, while comparing against data, the running may have significant 
impact and may be degenerate to some extent with $\fnl^{\rm type}$ in its
constraints.

For the templates studied so far, we find that the value of the respective 
amplitudes of non-Gaussianity $\fnl^{\rm type}$, that is required to produce 
discernible effect on the $C_\ell$s to be much higher than the existing 
bounds on them arrived at from direct comparison against data by Planck~\cite{Planck:2019kim}.
However, this exercise gives us insights on the computation of the integrals 
involved and helps us to employ the technique to $\fnl(k_1,k_2,k_3)$ with
non-trivial scale dependences as we shall discuss next.


\subsection{Oscillatory Type}\label{subsec:osc}

We turn to a rather interesting case of oscillatory features in the power 
spectrum and the bispectrum.
Such spectra are observed in models of inflation containing oscillatory features 
in potentials, sharp transitions during the field evolution or non-inflationary
initial epoch~\cite{Flauger:2010ja,Brandenberger:2012aj,Sreenath:2014nca,Ragavendra:2020old}.
The scalar power in this case is parametrized as 
\begin{equation}
\ps^{\rm osc}(k) = A_{_{\rm S}}\left(\frac{k}{k_\ast}\right)^{\ns-1}\,
\left\{1 + b\,\sin\left[\omega\ln\left(\frac{k}{k_o}\right)\right]\right\}\,,
\label{eq:ps-osc}
\end{equation}
where $b$ quantifies the strength and $\omega$ the frequency of the oscillatory 
feature. The scalar bispectrum is modelled as~\cite{Planck:2019kim}
\begin{equation}
{\cal B}^{\rm osc}(k_1,k_2,k_3) = \frac{6A_{_{\rm S}}^2\fnl^{\rm osc}}{(k_1k_2k_3)^2}
\sin\left[\omega\ln\left(\frac{k_1+k_2+k_3}{k_o}\right)\right]\,.
\end{equation}
Evidently the amplitude of the bispectrum and hence the associated $\fnl(k_1,k_2,k_3)$
is determined by $\fnl^{\rm osc}$. We should note that in the above expression 
the factor $A_{_{\rm S}}$ is included while ignoring the tilt and running of the
bispectrum. So, when we compute $\fnl(k_1,k_2,k_3)$, the tilt in the power spectra
in the denominator shall also be consistently ignored. In other words, if we 
include the tilt of $(k/k_\ast)^{(\ns-1)}$ in the parametrization of the bispectrum, then it
will be cancelled by corresponding term arising from the power spectra in the
denominator [cf.~Eq.~\eqref{eq:fnl-ps-g}]. Thus, the $\fnl(k_1,k_2,k_3)$ in this case
shall be
\begin{eqnarray}
\fnl(k_1,k_2,k_3) &=& -20\sqrt{2\pi}(k_1k_2k_3)A_{_{\rm S}}^2\fnl^{\rm osc}
\frac{\sin\left[\omega\ln\left(\frac{k_1+k_2+k_3}{k_o}\right)\right]}
{\left[k_1^3\ps^{\rm osc}(k_2)\ps^{\rm osc}(k_3) + k_2^3\ps^{\rm osc}(k_1)\ps^{\rm osc}(k_3) + k_3^3\ps^{\rm osc}(k_1)\ps^{\rm osc}(k_2)
\right]} \\
&=& -20\sqrt{2\pi}(k_1k_2k_3)\fnl^{\rm osc}
\frac{\sin\left[\omega\ln\left(\frac{k_1+k_2+k_3}{k_o}\right)\right]}
{\left\{k_1^3\,\left[1+b\sin\left(\omega\ln\frac{k_2}{k_o}\right)\right]
\left[1+b\sin\left(\omega\ln\frac{k_3}{k_o}\right)\right] + 
2~{\rm permutations}\right\}}\,.
\end{eqnarray}
Substituting $\ps^{\rm osc}(k)$ and $\fnl(k_1,k_2,k_3)$ above in 
Eq.~\eqref{eq:pc-fnl}, we obtain $\pc^{\rm osc}(k)$ of the form 
\begin{eqnarray}
\pc^{\rm osc}(k) &=& 
288\,\pi\,\,\bigg[A_{_{\rm S}}\left(\frac{k}{k_\ast}\right)^{\ns-1}
\fnl^{\rm osc}\bigg]^2
\int_0^\infty{\rm d}x \int^{1+x}_{\vert 1-x\vert}{\rm d}y\,xy
\left[1+b\sin\left(\omega \ln \f{kx}{k_o} \right)\right]
\left[1+b\sin\left(\omega \ln \f{ky}{k_o} \right)\right] \nn \\
& & \times \sin^2\left[ \omega \ln \left( \f{k}{k_o} (1+x+y) \right) \right]
\Bigg\{ \left[1+b\sin\left(\omega \ln \f{kx}{k_o} \right)\right]
\left[1+b\sin\left(\omega \ln \f{ky}{k_o} \right)\right] \nn \\
& & + x^3\,\left[1+b\sin\left(\omega \ln \f{k}{k_o} \right)\right]
\left[1+b\sin\left(\omega \ln \f{ky}{k_o} \right)\right]
+ y^3\,\left[1+b\sin\left(\omega \ln \f{k}{k_o} \right)\right]
\left[1+b\sin\left(\omega \ln \f{kx}{k_o} \right)\right] \Bigg\}^{-1}.\;\;\;\;\;\;
\label{eq:pc-osc-full}
\end{eqnarray}
Note that we have ignored the terms such as $x^{\ns-1}$ and $y^{\ns-1}$ as
in the cases before, since we are interested in the order of magnitude of the
above integral and these terms provide negligible contribution to this estimate.
Moreover, the integrand does not diverge at $(x,y)=(0,1)$ or $(1,0)$ due to the
well behaved shape of the bispectrum (cf. App.~\ref{app:integrands}).

To perform the double integral above, we shall utilize the fact that the 
parameter $b$ typically takes values $b<1$. So, we shall first focus
on the terms of order $b^0$ and then proceed to higher order terms in our 
calculation of $\pc^{\rm osc}(k)$. The $\fnl(k,kx,ky)$ as it appears in the
integral of $\pc^{\rm osc}(k)$, can then be expanded as
\begin{eqnarray}
\fnl(k,kx,ky) &=& -20\sqrt{2\pi}\,\fnl^{\rm osc}\frac{xy}{1+x^3+y^3}
\bigg[\sin\left(\omega\ln\frac{k}{k_o}\right) 
\cos\left[\omega \ln (1+x+y)\right] \nn \\
& & +
\cos \left(\omega\ln\frac{k}{k_o}\right) 
\sin \left[\omega \ln (1+x+y)\right] \bigg] + {\cal O}(b) + {\cal O}(b^2)\,.
\end{eqnarray}
We obtain the correction to the scalar power $\pc(k)$ at the leading order of 
${\cal O}(b^0)$ to be
\begin{eqnarray}
\pc^{\rm osc}(k) &\simeq & 2\pi\left[12\,A_{_{\rm S}}
\left(\frac{k}{k_\ast}\right)^{(\ns-1)}\fnl^{\rm osc}\right]^2
\int_0^\infty{\rm d}x \int^{1+x}_{\vert 1-x\vert}{\rm d}y
\frac{1}{(1+x^3+y^3)^2} \nn \\
& & \times
\bigg[\sin\left(\omega\ln\frac{k}{k_o}\right) 
\cos\left[\omega \ln (1+x+y)\right]
+ \cos \left(\omega\ln\frac{k}{k_o}\right) 
\sin \left[\omega \ln (1+x+y)\right] \bigg]^2\,.
\end{eqnarray}
On simplifying the integral further, we obtain
\begin{eqnarray}
\pc^{\rm osc}(k) &=& \pi\left[12\,A_{_{\rm S}}
\left(\frac{k}{k_\ast}\right)^{(\ns-1)}\fnl^{\rm osc}\right]^2
\bigg[ {\cal I}_1
- \cos\left(2\omega\ln\frac{k}{k_o}\right) {\cal I}_2
+ \sin\left(2\omega\ln\frac{k}{k_o}\right) {\cal I}_3  \bigg]\,,
\label{eq:pc-b0}
\end{eqnarray}
where the terms $\cal I$s are integrals whose explicit forms are provided 
in App.~\ref{app:integrands}. We know that the integrands of $\cal I$ are 
unity at maximum and hence the terms themselves evaluate to ${\cal O}(1)$ values. 
As can be expected, $\pc^{\rm osc}(k)$, at the leading order, has an oscillatory 
pattern of frequency $2\omega$, while the original spectrum $\ps^{\rm osc}(k)$ has 
a frequency of $\omega$. 

Now let us turn to the terms of ${\cal O}(b)$. The parameter $\fnl(k,kx,ky)$ 
expanded up to linear order in $b$ is 
\begin{eqnarray}
\fnl(k,kx,ky) &=& -20\sqrt{2\pi}\fnl^{\rm osc}\frac{xy}{(1+x^3+y^3)}
\sin\left[\omega\ln\left(\frac{k}{k_o}\right)+\omega\ln(1+x+y)\right] \nn\\
& & \times \Bigg\{1 - \frac{b}{(1+x^3+y^3)}
\Bigg[\sin\left[\omega\ln\left(\frac{k}{k_o}\right)+\omega\ln x\right]
+ \sin\left[\omega\ln\left(\frac{k}{k_o}\right)+\omega\ln y\right] \nn \\
& & + (x^3+y^3)\sin\left[\omega\ln\left(\frac{k}{k_o}\right)\right]
+ x^3\sin\left[\omega\ln\left(\frac{k}{k_o}\right)+\omega\ln y\right]
+ y^3\sin\left[\omega\ln\left(\frac{k}{k_o}\right)+\omega\ln x\right]\Bigg]
\Bigg\}.
\end{eqnarray}
Substituting this expression in $\pc(k)$ along with $\ps^{\rm osc}(k)$ terms
expanded to the same order in $b$, we obtain
\begin{eqnarray}
\pc^{\rm osc}(k) &=& 
288\,\pi\,\,\bigg[A_{_{\rm S}}\left(\frac{k}{k_\ast}\right)^{\ns-1}
\fnl^{\rm osc}\bigg]^2
\int_0^\infty{\rm d}x \int^{1+x}_{\vert 1-x\vert}{\rm d}y \f{1}{x^2y^2}
\Bigg[\frac{xy}{1+x^3+y^3}\sin\bigg[\omega\ln\left(\frac{k}{k_o}\right) + \omega\ln(1+x+y)\bigg]\Bigg]^2 \nn \\
& & \times \Bigg\{1 - \frac{b}{(1+x^3+y^3)}
\Bigg[\sin\left[\omega\ln\left(\frac{k}{k_o}\right)+\omega\ln x\right]
+ \sin\left[\omega\ln\left(\frac{k}{k_o}\right)+\omega\ln y\right] \nn \\
& & + (x^3+y^3)\sin\left[\omega\ln\left(\frac{k}{k_o}\right)\right]
+ x^3\sin\left[\omega\ln\left(\frac{k}{k_o}\right)+\omega\ln y\right]
+ y^3\sin\left[\omega\ln\left(\frac{k}{k_o}\right)+\omega\ln x\right]\Bigg]
\Bigg\}^2.
\end{eqnarray}
In this integral, we retain only terms up to ${\cal O}(b)$ and ignore terms with
${\cal O}(b^2)$ or higher order. We also ignore ${\cal O}(b^0)$ as they have been 
already taken care of. 
Grouping these terms and simplifying, we obtain $\pc^{\rm osc}(k)$ at the order 
of ${\cal O}(b)$ to be
\begin{eqnarray}
\pc^{\rm osc}(k) &\simeq & -\pi\, b\, \left[24\,A_{_{\rm S}}
\left(\frac{k}{k_\ast}\right)^{(\ns-1)}\fnl^{\rm osc}\right]^2
\bigg[ \sin^3\left( \omega\ln\frac{k}{k_o}\right) {\cal I}^{(b)}_1 
+ \sin^2\left( \omega\ln\frac{k}{k_o}\right)\cos\left( \omega\ln\frac{k}{k_o}\right){\cal I}^{(b)}_2 \nn \\
& & + \sin\left( \omega\ln\frac{k}{k_o}\right)\cos^2\left( \omega\ln\frac{k}{k_o}\right){\cal I}^{(b)}_3
+ \cos^3\left( \omega\ln\frac{k}{k_o}\right){\cal I}^{(b)}_4
+ \sin^2\left( \omega\ln\frac{k}{k_o}\right)\cos\left( \omega\ln\frac{k}{k_o}\right){\cal I}^{(b)}_5 \nn \\
& & + \sin\left( \omega\ln\frac{k}{k_o}\right)\cos^2\left( \omega\ln\frac{k}{k_o}\right){\cal I}^{(b)}_6 \bigg]\,.
\label{eq:pc-b1}
\end{eqnarray}
Here the integrals ${\cal I}^{(b)}$s are again ${\cal O}(1)$ quantities whose 
exact expressions are given in App.~\ref{app:integrands}.
We find that the corrections of ${\cal O}(b)$ are of frequency $3\omega$.
The total $\pc^{\rm osc}(k)$ shall be the sum of the terms obtained in 
Eqs.~\eqref{eq:pc-b0} and~\eqref{eq:pc-b1}.
We restrict our calculation up to ${\cal O}(b)$ and compute the angular spectrum
arising due to this $\pc^{\rm osc}(k)$. Since, the quantities $\cal I$s are
${\cal O}(1)$ we set their values to unity in our computation of $C_\ell$s.
This approximation is done just to understand the general behavior and parametric 
dependences of $C_\ell$s in this case. One should perform these integrals
exactly while comparing against the data to arrive at constraints on the
parameters.

\begin{figure}[t]
\centering
\includegraphics[width=8cm,height=6cm]{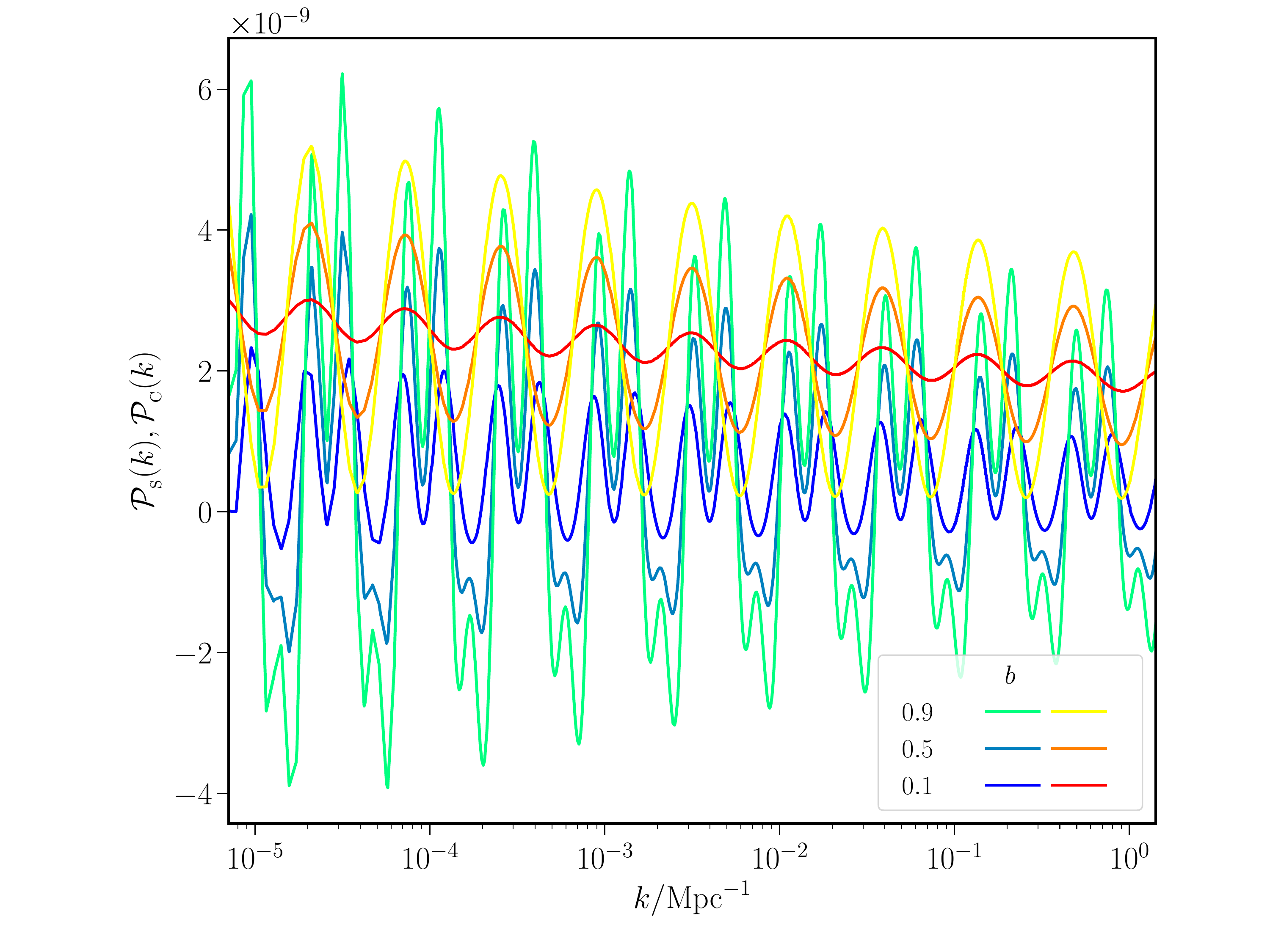}
\includegraphics[width=8cm,height=6cm]{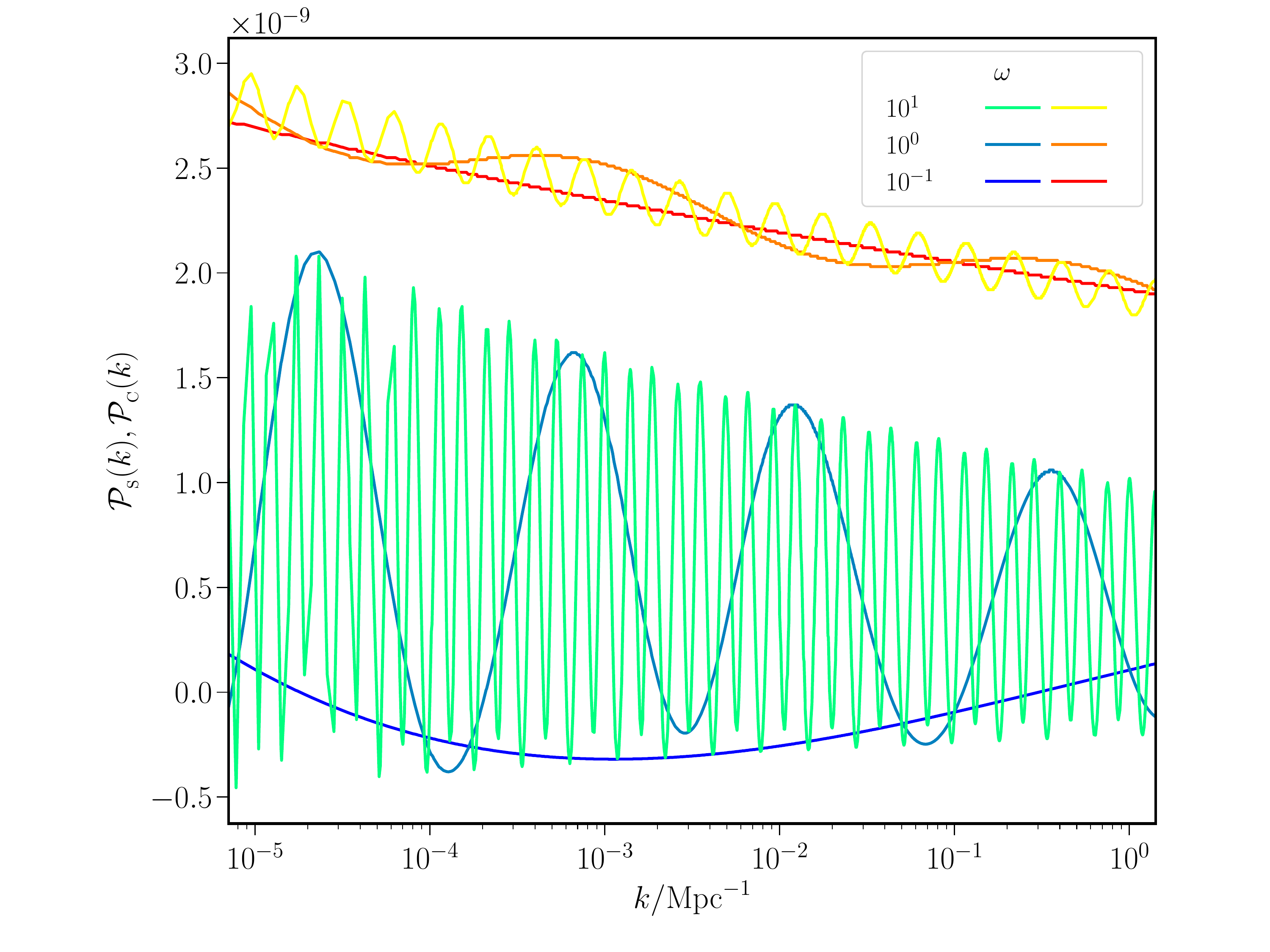}
\vskip -0.1in
\caption{The $\ps^{\rm osc}(k)$ (in shades of red to yellow) and $\pc^{\rm osc}(k)$
(in shades of blue to green) are presented for different values of $b$ (on left) 
and $\omega$ (on right). 
Note that in the expression of $\pc^{\rm osc}(k)$, $b$ determines the strength
of oscillations of the component that has frequency of $3\omega$. Hence, increase 
in $b$ enhances this part of $\pc^{\rm osc}(k)$ while the contribution with 
frequency of $2\omega$ remains unaffected in amplitude. We set $\omega=5$ to 
arrive at this behavior.
On the other hand, as $\omega$ is increased, the oscillatory patterns are more 
pronounced in $\pc(k)$ than in $\ps(k)$. We set $b=0.05$ to obtain this plot.
The other related parameters are set to be $\fnl^{\rm osc}=500$, and 
$k_o/\,\mpcinv=10^{-1}$ in obtaining these plots, so that the features of interest
are better illustrated.}
\label{fig:PsPc_osc}
\end{figure}
In Fig.~\ref{fig:PsPc_osc} we show the variation of $\ps^{\rm osc}(k)$ and $\pc^{\rm osc}(k)$ for a range of parameters $b$ and $\omega$. The parameter $b$, in the expression of $\pc^{\rm osc}(k)$, determines the strength of oscillations of the component that has frequency of $3\omega$. Hence, for a given value of $\omega$ increasing $b$ enhances this part of $\pc^{\rm osc}(k)$ while the dominant contribution with frequency of $2\omega$ remains unaffected in amplitude. On the other hand, if we fix the value of $b$ and increase $\omega$, we observe that the oscillatory patterns are more pronounced in $\pc^{\rm osc}(k)$ than in $\ps^{\rm osc}(k)$. For both the plots, we have set $\fnl^{\rm osc}=500$, and 
$k_o/\,\mpcinv=10^{-1}$.

\begin{figure}[t]
\centering
\includegraphics[width=12cm,height=8cm]{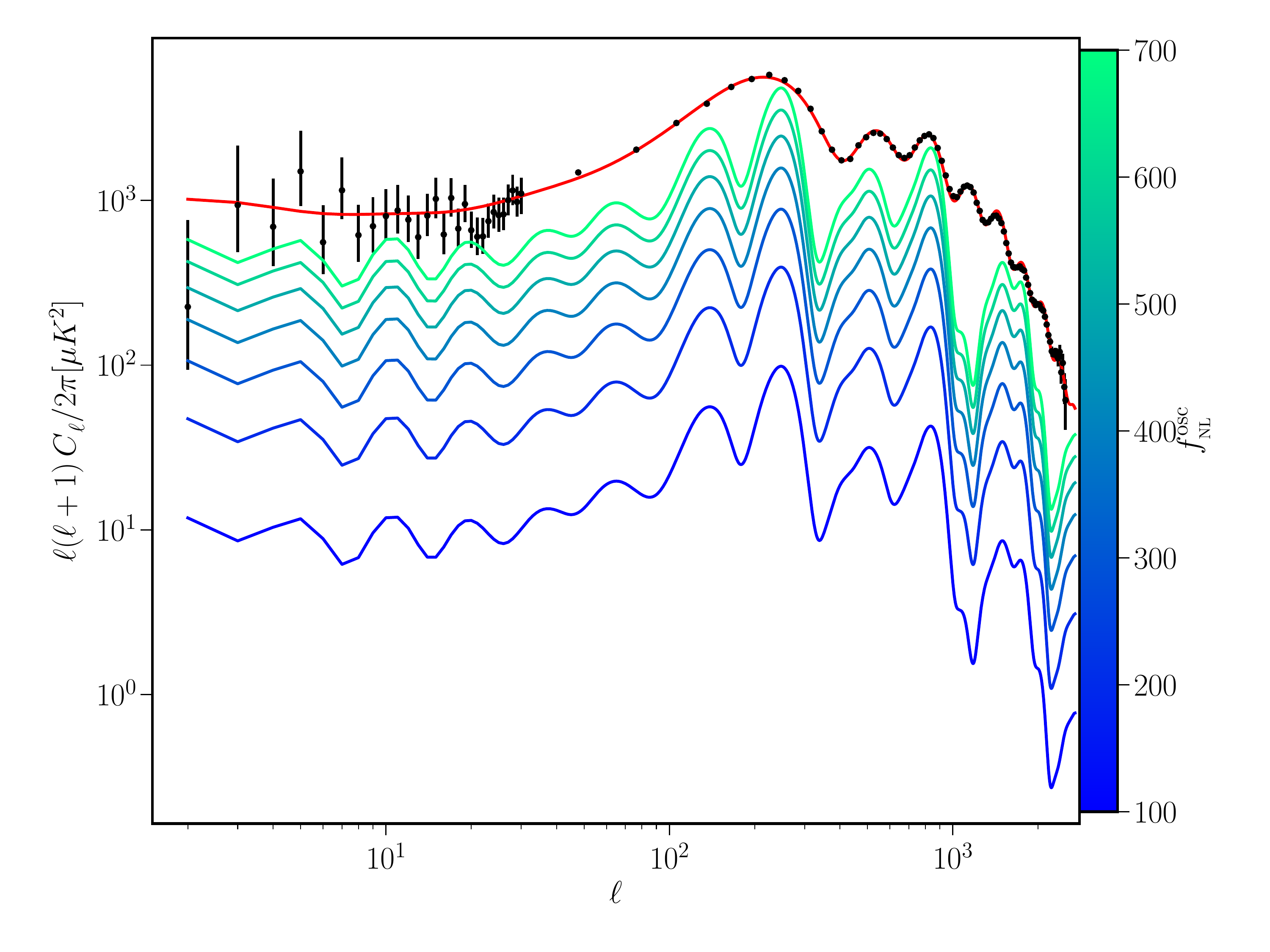}
\vskip -0.1in
\caption{The angular spectrum due to $\pc^{\rm osc}(k)$ is presented (in shades 
of blue to green) across a range of the parameter $\fnl^{\rm osc}$ along with
the spectrum due to $\ps^{\rm osc}(k)$ (in red) and the data points (in black).
As expected, the amplitude of $C_{\ell}$s due to $\pc^{\rm osc}(k)$ increases as 
this parameter is increased. However, the crucial point to note is that the range
of $\fnl^{\rm osc}$ required to get amplitudes comparable to the standard 
$C_{\ell}$s is just $\sim 500$, which is realizable in models in the literature
that produce such oscillatory bispectra.
We have fixed other related parameters to be $b=5 \times 10^{-2}, \omega=5$ and 
$k_o/\,\mpcinv=10^{-1}$ in obtaining these plots.}
\label{fig:ClsTT_osc_fnlosc}
\end{figure}
\begin{figure}[h]
\centering
\includegraphics[width=12cm,height=8cm]{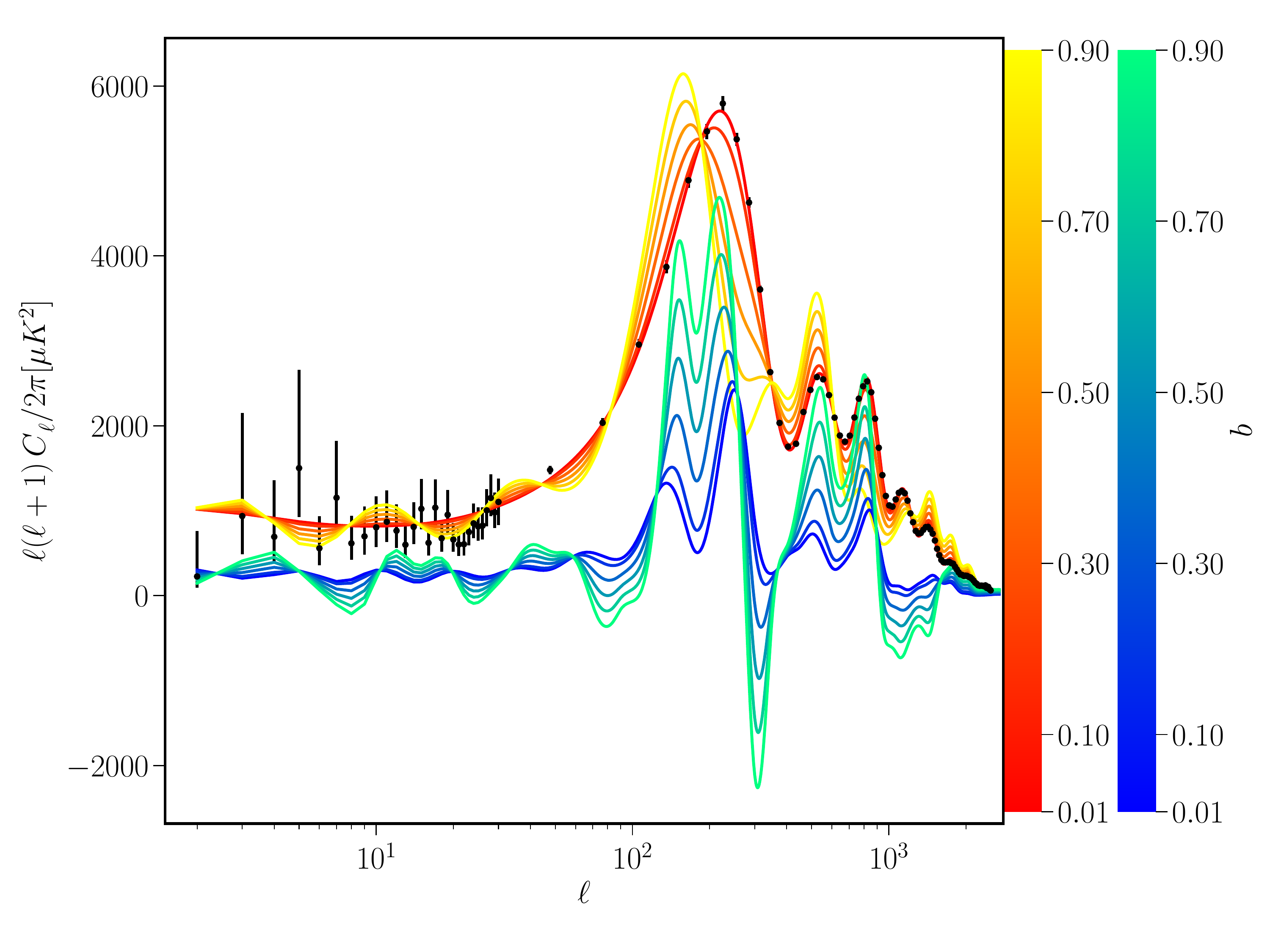}
\vskip -0.1in
\caption{The angular spectra due to $\pc^{\rm osc}(k)$ and $\ps^{\rm osc}(k)$ 
are presented (in shades of blue to green, and red to yellow respectively),
across a range of the parameter $b$ which determines the strength of 
oscillations in $\ps^{\rm osc}(k)$. Note that in the expression of 
$\pc^{\rm osc}(k)$, $b$ determines the strength of oscillations of the component
that has frequency of $3\omega$. Hence, variation of $b$ affects this part of
$\pc^{\rm osc}(k)$ while the contribution with frequency of $2\omega$ still
remains unaffected in amplitude. The other related parameters are set to be 
$\fnl^{\rm osc}=500, \omega=5$ and $k_o/\,\mpcinv=10^{-1}$ in this figure.}
\label{fig:ClsTT_osc_b}
\end{figure}
\begin{figure}[h]
\centering
\includegraphics[width=12cm,height=8cm]{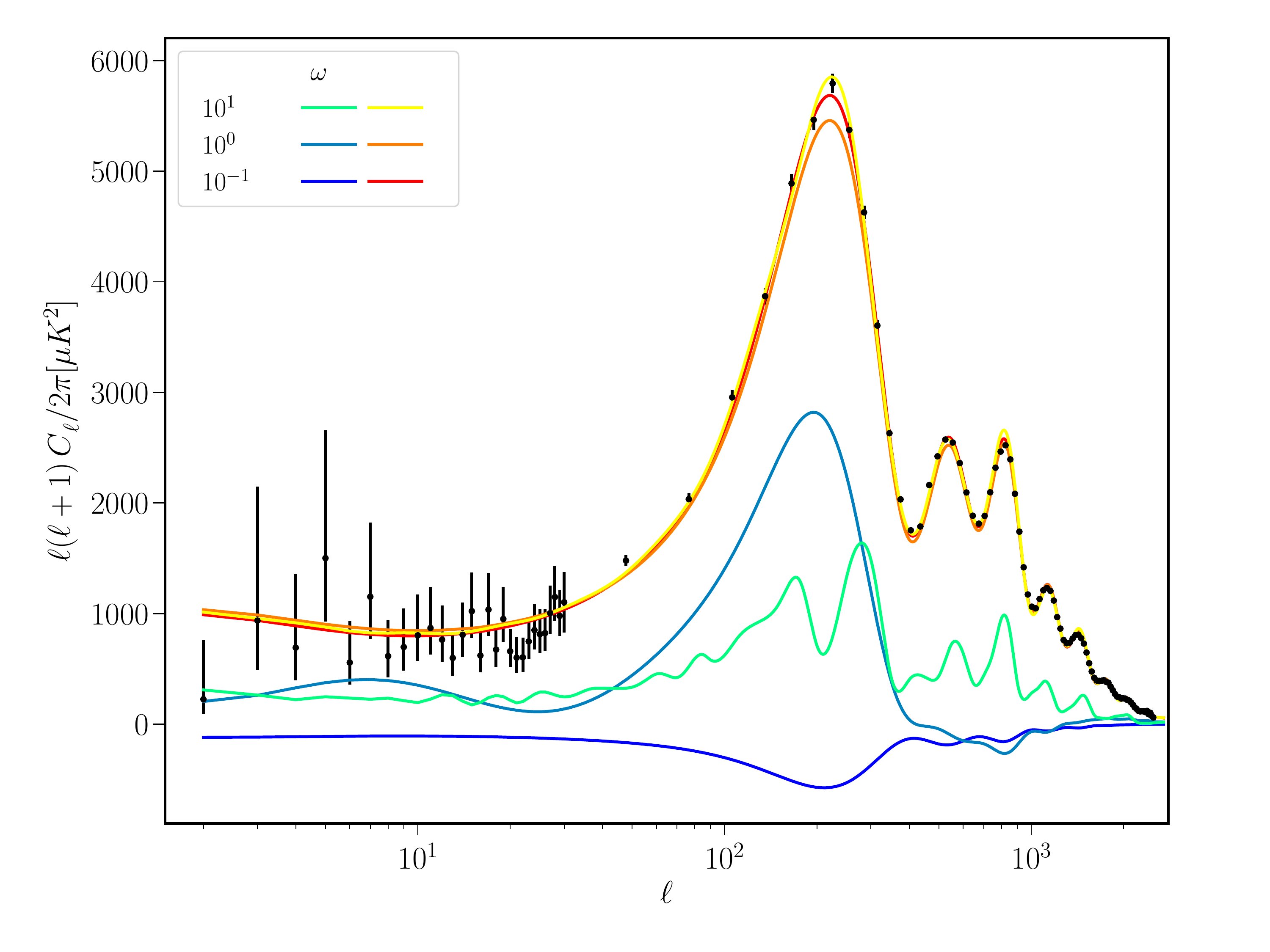}
\vskip -0.1in
\caption{The angular spectra due to $\pc^{\rm osc}(k)$ and $\ps^{\rm osc}(k)$ 
are presented (in shades of blue to green, and red to yellow respectively)
across a range of the parameter $\omega$ which determines the frequency of 
oscillations in the power and the bi-spectra in this template. As the frequency 
is increased, the oscillatory patterns are more pronounced in $\pc(k)$ than in 
$\ps(k)$. We have set $\fnl^{\rm osc}=500, b=5\times 10^{-2}$ and 
$k_o/\,\mpcinv=10^{-1}$ in obtaining these plots.}
\label{fig:ClsTT_osc_omega}
\end{figure}
We illustrate the CMB angular spectra arising in this case in
Figs.~\ref{fig:ClsTT_osc_fnlosc},~\ref{fig:ClsTT_osc_b} 
and~\ref{fig:ClsTT_osc_omega}. Note that along with the $C_\ell$s due to 
$\pc^{\rm osc}(k)$, we plot the original $C_\ell$s in this case using 
$\ps^{\rm osc}(k)$ as given in~\eqref{eq:ps-osc}. We set the default values of 
the parameters to be $b=5\times 10^{-2},\,\fnl^{\rm osc} = 500,\,\omega = 5,
k_o/\,\mpcinv = 0.1$ about which we vary them individually to illustrate our results. 
As we have observed in previous cases, the amplitude of $C_\ell$s due to 
$\pc^{\rm osc}(k)$ is directly proportional to the strength of the non-Gaussianity 
parameter, which in this case is $\fnl^{\rm osc}$. 
This is clearly observed in Fig.~\ref{fig:ClsTT_osc_fnlosc}. 
A crucial point to note is that the value of $\fnl^{\rm osc}$ required for 
$\pc^{\rm osc}(k)$ to leave imprints on the CMB spectrum is around $500$. 
Such values of $\fnl^{\rm osc}$ is achievable in realistic scenarios of 
inflation~\cite{Flauger:2009ab,Hazra:2014goa,Ragavendra:2020old}. 
Therefore, this method shall be efficient in obtaining constraints on 
$\fnl(k_1,k_2,k_3)$ arising from these models, in the absence of strict 
direct constraint from data.

The strength of the feature in $\ps^{\rm osc}(k)$, determined by $b$ also has 
an interesting effect on $C_\ell$s, as presented in Fig.~\ref{fig:ClsTT_osc_b}.
Recall that $b$ is typically less than unity and so we vary it between 
$10^{-2}$ and $0.9$. We find that the oscillatory patterns are more pronounced 
in both the standard $C_\ell$s and the $C_\ell$s due to $\pc^{\rm osc}(k)$. But, 
as can be seen in Eq.~\eqref{eq:pc-b0}, the part of contribution to 
$\pc^{\rm osc}(k)$ whose oscillatory features are of frequency $2\omega$ is 
independent of $b$. On the other hand, as given in Eq.~\eqref{eq:pc-b1}, 
the part of $\pc^{\rm osc}(k)$ with frequency $3\omega$ is linearly
proportional to $b$. Hence, variation in $b$ increases the contribution to 
$C_\ell$s due $\pc^{\rm osc}(k)$ with $3\omega$ oscillations in wavenumbers.
Lastly, varying the frequency of the feature in the spectra $\omega$, directly 
the varies frequency of oscillatory pattern in $C_\ell$s. However, yet another
interesting point to note is that, while the standard $C_\ell$s due to 
$\ps^{\rm osc}(k)$ receive contribution with frequency of $\omega$, the 
$C_\ell$s due to $\pc^{\rm osc}(k)$ receive contributions containing $2\omega$ 
and $3\omega$. 
This induces a pronounced difference in the behavior of $C_\ell$s as $\omega$ is 
varied, as observed in Fig.~\ref{fig:ClsTT_osc_omega}.
Therefore, the different effects that $b$ and $\omega$ have on the angular 
spectrum through $\pc^{\rm osc}(k)$ suggest that the constraints on them arising
from $\ps^{\rm osc}(k)$ alone shall get modified when including $\pc^{\rm osc}(k)$ 
while comparing against the data.


\section{Starobinsky model}\label{sec:staro}

As a final illustration of our method, we employ it for a realistic model of
inflation driven by a potential that was originally proposed by 
Starobinsky~\cite{Starobinsky:1992ts}.
This model is interesting in that it has a sudden change in the slope of the 
potential and hence induces interesting features in the power spectrum thereby 
improving the fit to the data. Moreover, there also arise non-trivial features
in the scalar bispectrum~\cite{Martin:2011sn,Martin:2014kja,Ragavendra:2020old}.
The form of the potential in this model is given by~\cite{Starobinsky:1992ts}
\begin{eqnarray}
    V(\phi) &=& 
\left\{
    \begin{array}{lr}
        V_0 + A_+ (\phi - \phi_0), & \text{for } \phi > \phi_0\,,\\
        V_0 + A_- (\phi - \phi_0), & \text{for } \phi < \phi_0\,,
    \end{array}
\right.
\end{eqnarray}
where $\phi_0$ is the point where the slope changes from $A_+$ to $A_-$. $V_0$ is 
the value of potential at $\phi = \phi_0$. 
Let us briefly discuss the relevant quantities governing the 
evolution of the field in this model, before studying the power and bi-spectra.

If we consider $V_0$ in potential to be dominant, then the first slow roll 
parameter, $\epsilon_1 \equiv -\dot{H}/H^2$, remains smaller than unity 
throughout the evolution of the field. We can write the first slow roll 
parameter $\epsilon_1$, over the two regimes of evolution as 
\begin{eqnarray}
    \epsilon_{1+} &\simeq & \f{A_+^2}{18 M_{\rm Pl}^{2} H_0^4} \text{,}\\
    \epsilon_{1-} &\simeq & \f{A_-^2}{18 M_{\rm Pl}^{2} H_0^4}\left[1 - \f{\Delta A}{A_-}\left(\f{\eta}{\eta_0}\right)^3\right]^2 \text{,}
\end{eqnarray}
where $H_0 \simeq \sqrt{V_0/(3 M_{\rm Pl}^{2})}$, $\Delta A = A_- - A_+$ and $\eta_0$ 
denotes the conformal time at which $\phi=\phi_0$. The second slow roll parameter 
$\epsilon_2 \equiv d \ln{\epsilon_1}/dN$ over the two regimes are
\begin{eqnarray}
    \epsilon_{2+} &\simeq & 4 \epsilon_{1+}\,, \\
    \epsilon_{2-} &\simeq & \f{6\Delta A}{A_-}\f{\left(\f{\eta}{\eta_0}\right)^3}{1 - \f{\Delta A}{A_-}\left(\f{\eta}{\eta_0}\right)^3} + 4\epsilon_{1-}\text{.}
\end{eqnarray}
At transition, $\epsilon_{2-}$ becomes large and so does the time 
derivative of $\epsilon_{2-}$.
This sharp behavior can be approximated by a Dirac delta function (see Ref. \cite{Martin:2014kja})
\begin{eqnarray}
\epsilon_{2-}^{\prime} &\simeq & \f{6\Delta A\, \eta_0}{A_+\, \eta} \delta^{(1)}(\eta - \eta_0)\,.
\end{eqnarray}

Utilizing these behaviors of the background quantities, we can solve the
scalar perturbations and obtain the analytical expression for the power spectrum 
to be
\begin{eqnarray}
    \ps(k) &=&  \ps^0 |\alpha_{k} - \beta_{k}|^2\,,
\end{eqnarray}
where 
\begin{eqnarray}
\ps^0 &=& \f{1}{12\pi^2}\left(\f{V_0}{M_{\rm Pl}^{4}}\right)\left(\f{V_0}{A_- M_{\rm Pl}}\right)^2.
\end{eqnarray}
The functions $\alpha_{k}$ and $\beta_{k}$ are the Bogoliubov coefficients obtained
by matching the mode functions of perturbations before and after transition at $\phi_0$.
They are given in terms of model parameters as
\begin{eqnarray}
    \alpha_{k} &=& 1 + i\f{3\Delta A}{2A_+}\f{k_0}{k}\left(1+\f{k_0^2}{k^2}\right), \\
    \beta_{k} &=& \f{3\Delta A}{2A_+}\f{k_0}{k}\left(\sin \left(\f{2 k}{k_0}\right)-\f{k_0^2}{k^2}\sin \left(\f{2 k}{k_0}\right)+2\f{k_0}{k}\cos \left(\f{2 k}{k_0}\right)\right) \nn \\
    & & -i\,\f{3\Delta A}{2A_+}\f{k_0}{k}\left(\cos \left(\f{2 k}{k_0}\right)-\f{k_0^2}{k^2}\cos \left(\f{2 k}{k_0}\right)-2\f{k_0}{k}\sin \left(\f{2 k}{k_0}\right)\right)\,,
\end{eqnarray}
where $k_0 = -1/\eta_0$ represents the mode that leaves the Hubble radius at transition. Thus complete expression of $\ps(k)$ is~\cite{Martin:2011sn}
\begin{eqnarray}
\ps(k) &=&  \f{1}{12\pi^2}\left(\f{V_0}{M_{\rm Pl}^{4}}\right)\left(\f{V_0}{A_- M_{\rm Pl}}\right)^2 \left\{1-\f{3 \Delta A}{A_+}\f{k_0}{k}\left[\left(1-\f{k_0^2}{k^2}\right)\sin \left(\f{2k}{k_0}\right) + \f{2k_0}{k}\cos \left(\f{2k}{k_0}\right)\right]\right. \nn \\
    & & + \left.\f{9 \Delta A^2}{2 A_+^2}\f{k_0^2}{k^2}\left(1+\f{k_0^2}{k^2}\right)\left[1+\f{k_0^2}{k^2} - \f{2k_0}{k}\sin \left(\f{2k}{k_0}\right) + \left(1-\f{k_0^2}{k^2}\right)\cos \left(\f{2k}{k_0}\right) \right]\right\}.
    \label{eq:ps_staro}
\end{eqnarray}

As to the bispectrum ${\cal B}(k_1,k_2,k_3)$, there are typically nine terms
that capture contributions from the cubic order action of the scalar 
perturbations\,\footnote{These terms are denoted as $G_i(k_1,k_2,k_3)$ where
$i=\{1,...9\}$ and they arise from six bulk terms and three boundary terms of the
cubic order action. The total contribution $G(k_1,k_2,k_3)$ is related to the 
bispectrum that we denote as ${\cal B}(k_1,k_2,k_3)$ through a simple numerical 
factor as ${\cal B}(k_1,k_2,k_3) = (2\pi)^{-9/2}G(k_1,k_2,k_3)$. For detailed 
discussion regarding each of these terms, see 
Refs.~\cite{Maldacena:2002vr,Martin:2011sn,Arroja:2011yj,Ragavendra:2020old,
Ragavendra:2023ret}.}. In this model, the dominant terms turn out to be 
$G_4(k_1,k_2,k_3) + G_7(k_1,k_2,k_3)$ as $G_4(k_1,k_2,k_3)$ contains the 
quantity $\epsilon_1\epsilon_2^{\prime}$.
$G_7(k_1,k_2,k_3)$, which arises due to a boundary term in the action and is 
typically absorbed through field redefinition, complements the super-Hubble
contribution of $G_4(k_1,k_2,k_3)$~\cite{Martin:2011sn,Hazra:2012yn}. The 
expression for $G_4(k_1,k_2,k_3)$ is
\begin{eqnarray}
    G_4(k_1,k_2,k_3) &=& \left[f_{k_1}(\eta_e)f_{k_2}(\eta_e)f_{k_3}(\eta_e)\right]\mathcal{G}_4(k_1,k_2,k_3) + \left[f_{k_1}^*(\eta_e)f_{k_2}^*(\eta_e)f_{k_3}^*(\eta_e)\right]\mathcal{G}_4^*(k_1,k_2,k_3),
\end{eqnarray}
where $f_k(\eta)$ is the mode function and a prime denotes derivative 
with respect to the conformal time $\eta$. The time $\eta_e$ is conformal time
close to the end of inflation where the spectra are evaluated. 
The term ${\cal G}_4$ is the integral over $\eta$ arising from the cubic order
action term whose form is given by
\begin{eqnarray}
    \mathcal{G}_4(k_1,k_2,k_3) &=& i\int_{-k_0^{-1}}^{0} d\eta a^2 \epsilon_{1-}\epsilon_{2-}^{\prime} \left(f_{k_1}^*f_{k_2}^*f_{k_3}^{*\prime}+f_{k_1}^*f_{k_2}^{*\prime}f_{k_3}^*+f_{k_1}^{*\prime}f_{k_2}^*f_{k_3}^*\right).
\end{eqnarray}
The calculation thus far closely follows Refs.~\cite{Martin:2011sn,Martin:2014kja},
but without focusing on a specific limit of the configuration of the bispectrum
such as equilateral or squeezed limit. Proceeding further, to compute $\pc(k)$, we
shall utilize these complete expressions of power and bi-spectra.
If we rewrite the arguments of $G_4$ as $k, kx$ and $ky$ as required for
$\pc(k)$, and proceed to substitute and utilize the behaviors of the slow roll 
parameters, we obtain
\begin{eqnarray}
    G_4(k,kx,ky) &=& \Mpl^2\left[f_{k}(\eta_e)f_{kx}(\eta_e)f_{ky}(\eta_e)\right]\left(i\int_{-k_0^{-1}}^{0} d\eta a^2 \epsilon_{1-}\epsilon_{2-}^{\prime} \left(f_{k}^*f_{kx}^*f_{ky}^{*\prime}+f_{k}^*f_{kx}^{*\prime}f_{ky}^*+f_{k}^{*\prime}f_{kx}^*f_{ky}^*\right)\right) \nn \\
    & & +~{\rm complex~conjugate}\,, \\
    &=& \f{i}{H_0^2}\int_{-k_0^{-1}}^{0} \f{d\eta}{\eta^2} \f{A_-^2}{18H_0^4}\left[1 - \f{\Delta A}{A_-}\left(\f{\eta}{\eta_0}\right)^3\right]^2 \f{6\Delta A}{A_+}\f{\eta_0}{\eta}\delta^{(1)}(\eta - \eta_0) \nn \\
    & & \times \bigg[f_{k}(\eta_e)f_{kx}(\eta_e)f_{ky}(\eta_e)\left(f_{k}^*f_{kx}^*f_{ky}^{*\prime}+f_{k}^*f_{kx}^{*\prime}f_{ky}^*+f_{k}^{*\prime}f_{kx}^*f_{ky}^*\right) \nn \\
    & &- f_{k}^*(\eta_e)f_{kx}^*(\eta_e)f_{ky}^*(\eta_e)\left(f_{k}f_{kx}f_{ky}^{\prime}+f_{k}f_{kx}^{\prime}f_{ky}+f_{k}^{\prime}f_{kx}f_{ky}\right)\bigg]\,, \\
    &=& \f{i \Delta A A_+}{3 H_0^6 \eta_0^2} \bigg[f_{k}(\eta_e)f_{kx}(\eta_e)f_{ky}(\eta_e)\left(f_{k}^*f_{kx}^*f_{ky}^{*\prime}+f_{k}^*f_{kx}^{*\prime}f_{ky}^*+f_{k}^{*\prime}f_{kx}^*f_{ky}^*\right) \nn \\
    & & - f_{k}^*(\eta_e)f_{kx}^*(\eta_e)f_{ky}^*(\eta_e)\left(f_{k}f_{kx}f_{ky}^{\prime}+f_{k}f_{kx}^{\prime}f_{ky}+f_{k}^{\prime}f_{kx}f_{ky}\right)\bigg]_{\eta = \eta_0}.
\end{eqnarray}
The exact expressions of the mode functions and their derivatives evaluated at
$\eta_0$ as appearing in the above expression are
\begin{eqnarray}
    f_k(\eta) &=& \f{iH_0}{2\Mpl \sqrt{k^3\epsilon_{1+}}}\left(1-\f{ik}{k_0}\right)e^{\f{ik}{k_0}}\,, \\
    f_k^{\prime}(\eta) &\simeq & \f{-iH_0}{2\Mpl \sqrt{k^3\epsilon_{1+}}}\f{k^2}{k_0}e^{\f{ik}{k_0}}\,.
\end{eqnarray}
Substituting the complete forms of $f_k(\eta_e),f_k(\eta_0)$ and
$f^\prime_k(\eta_0)$ in $G_4(k,kx,ky)$ and simplifying we obtain
\begin{eqnarray}
 k^6 x^3 y^3 G_4(k,kx,ky) &=& F(k) \times Z(k,x,y) \nn \\
    &=& F(k) \times (-i)\bigg\{(\alpha_k - \beta_k)(\alpha_{kx} - \beta_{kx})(\alpha_{ky} - \beta_{ky})e^{\f{-ik(1+x+y)}{k_0}} 
    \bigg[\left(1+\f{ikx}{k_0}\right)\left(1+\f{iky}{k_0}\right)
    \nn \\
    & & + \left(1+\f{ik}{k_0}\right)\left(1+\f{ikx}{k_0}\right)y^2 + \left(1+\f{iky}{k_0}\right)\left(1+\f{ik}{k_0}\right)x^2 \bigg]
    -~{\rm complex~conjugate}\bigg\}\,,
\end{eqnarray}
where $F(k)$ is the factor determining the amplitude of $G_4$ and $Z(k,x,y)$ is 
the function responsible for scale dependence of the dimensionless quantity 
$k^6 x^3 y^3 G_4(k,kx,ky)$. The form of $Z(k,x,y)$ can be easily seen from above
whereas the function $F(k)$ is given by
\begin{eqnarray}
    F(k) &=& 6\pi^4\,\left(\ps^0\right)^2\f{k_0}{k}\f{\Delta A}{A_+}\left(1 + \f{\Delta A}{A_+}\right)\,.
\end{eqnarray}
On the other hand, the expression for $G_7(k_1,k_2,k_3)$ is
\begin{eqnarray}
    G_7(k_1,k_2,k_3) &=& \f{\epsilon_{2-}(\eta_e)}{2}(|f_{k_2}(\eta_e)|^2|f_{k_3}(\eta_e)|^2 + |f_{k_1}(\eta_e)|^2|f_{k_2}(\eta_e)|^2 + |f_{k_1}(\eta_e)|^2|f_{k_3}(\eta_e)|^2) \nn \\
    &\simeq & \f{A_-^2}{9H_0^4}\f{4\pi^4}{k_1^3 k_2^3k_3^3}\left(k_1^3\ps(k_2)\ps(k_3) + k_2^3\ps(k_1)\ps(k_3) + k_3^3\ps(k_1)\ps(k_2)\right)\,.
\end{eqnarray}
We should also mention that, in this model, due to the sudden change of slope in 
potential at $\phi_0$, one can observe an uncontrolled growth of $G_4(k,kx,ky)$
over the range of wavenumbers. 
To mitigate this growth, we have, by hand, introduced a term 
$1/[1 + (k+kx+ky)/k_{\rm reg}]$, in $G_4(k,kx,ky)$ which captures the effect of 
smoothening of the potential (see~Ref.~\cite{Martin:2014kja} for dedicated discussion 
regarding sharp transition and smoothening). We have set $k_{\rm reg} = 100\,k_0$ 
in our computation of $\pc(k)$.

We proceed to compute the $\pc(k)$ arising in this model due to
$G_4 (k,kx,ky)$ and $G_7 (k,kx,ky)$ and obtain
\begin{eqnarray}
    \pc(k) &=& \f{4}{(2\pi)^8} \int_0^\infty \d x \int_{\vert 1-x \vert}^{1 + x} \d y \f{\ps(kx)}{x^2}\,\f{\ps(ky)}{y^2} \,\bigg(\f{k^6x^3y^3G_4(k,kx,ky)}{\ps(kx)\ps(ky) + y^3\ps(k)\ps(kx) + x^3\ps(k)\ps(ky)} \nn \\
    & & + \f{4\pi^4 A_-^2 M_{\rm Pl}^{2}}{V_0^2} \bigg)^2.
    \label{eq:pc_staro}
\end{eqnarray}
Note that the $k$,$x$,$y$ dependent terms in the integrand due to $G_7$ are 
cancelled by the combination of power spectra that appear in the expression of
$\fnl(k,kx,ky)$ in the denominator [cf.~Eq.~\eqref{eq:fnl-ps-g}], giving us just a
constant proportional to $A_-/V_0$.
Further, we can neglect this term due to $G_7 (k,kx,ky)$ as $G_7(k,kx,ky)\ll 
G_4 (k,kx,ky)$. Using the exact expression $G_4(k,kx,ky)$, we see that the 
integrand of $\pc(k)$, as presented below, is quite non-trivial. 
\begin{eqnarray}
    \pc(k) &\simeq & \f{9}{16} \f{k_0^2}{k^2}\left(\f{A_-}{A_+}\right)^2\left(1 - \f{A_-}{A_+}\right)^2\left({\ps^0}\right)^2 \int_0^\infty \d x \int_{\vert 1-x \vert}^{1 + x} \d y \f{ |\alpha_{kx} - \beta_{kx}|^2}{x^2}\,\f{|\alpha_{ky} - \beta_{ky}|^2}{y^2} \nn \\
    & & \times \left(\f{Z(k,x,y)}{|\alpha_{kx} - \beta_{kx}|^2 |\alpha_{ky} - \beta_{ky}|^2 + y^3 |\alpha_{k} - \beta_{k}|^2 |\alpha_{kx} - \beta_{kx}|^2 + x^3 |\alpha_{k} - \beta_{k}|^2 |\alpha_{ky} - \beta_{ky}|^2}\right)^2.
    \label{eq:pc_staro2}
\end{eqnarray}
The amplitude of $\pc(k)$ is mainly determined by $\left(\ps^0\right)^2$ and 
${A_-}/{A_+}$.
We shall compute this $\pc(k)$ numerically due to the complicated nature of the 
integrand. Besides, just like the case in $\pc^{\rm loc}(k)$, the integrand 
exhibits a divergence at $(x,y)= (0,1),(1,0)$ (cf. Fig.~\ref{fig:Integrand_staro} 
in App.~\ref{app:integrands}). So, we choose the value of 
$k_{\rm min}/\,\mpcinv = 10^{-6}$ to regulate this divergence as done earlier. 
We have also set $k_{\rm max}/\,\mpcinv = 10^{2}$ for numerical evaluation.

\begin{figure}[t]
\centering
\includegraphics[width=12cm,height=8cm]{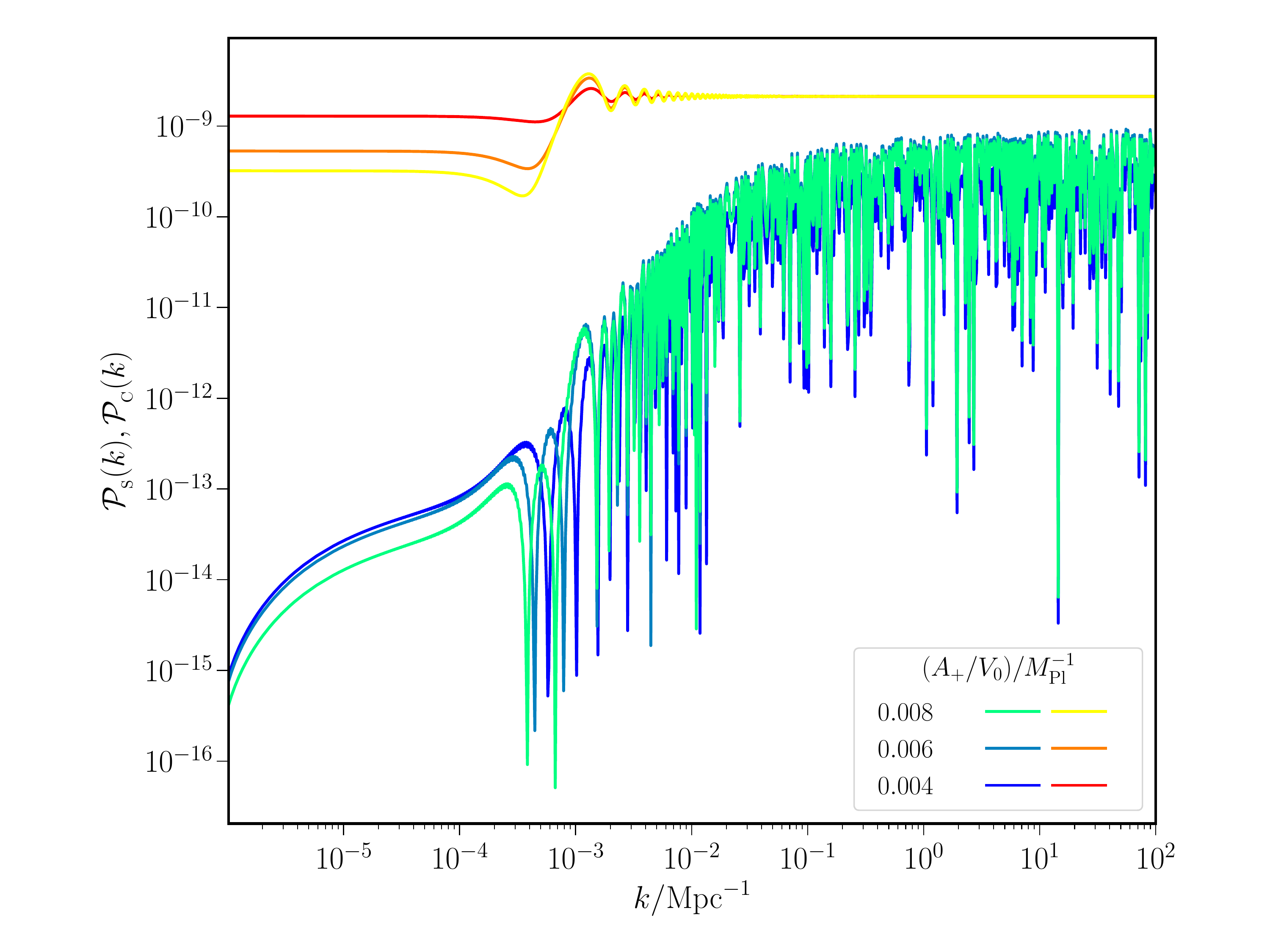}
\vskip -0.1in
\caption{We present $\ps(k)$ (in shades of red to yellow) and $\pc(k)$ (in shades of blue to green) arising from Starobinsky model for a range of the parameter $(A_+/V_0)/\Mpl^{-1}$. We have set $k_0/\,\mpcinv = 3.89 \times 10^{-4}$, $V_0/\Mpl^4 =2.48 \times 10^{-12}$ and $(A_-/V_0)/\Mpl^{-1}=3.14 \times 10^{-3}$ in obtaining these plots. As this parameter value is increased, the oscillatory patterns are more pronounced in $\ps(k)$ as expected. However, for $\pc (k)$, it is interesting to note that, the highest amplitude is achieved when $(A_+/V_0)/\Mpl^{-1}$ is twice that of $(A_-/V_0)/\Mpl^{-1}$ especially over large $k$ range. 
}
\label{fig:PsPc_StaroII_Ap}
\end{figure}
\begin{figure}[h]
\centering
\includegraphics[width=12cm,height=8cm]{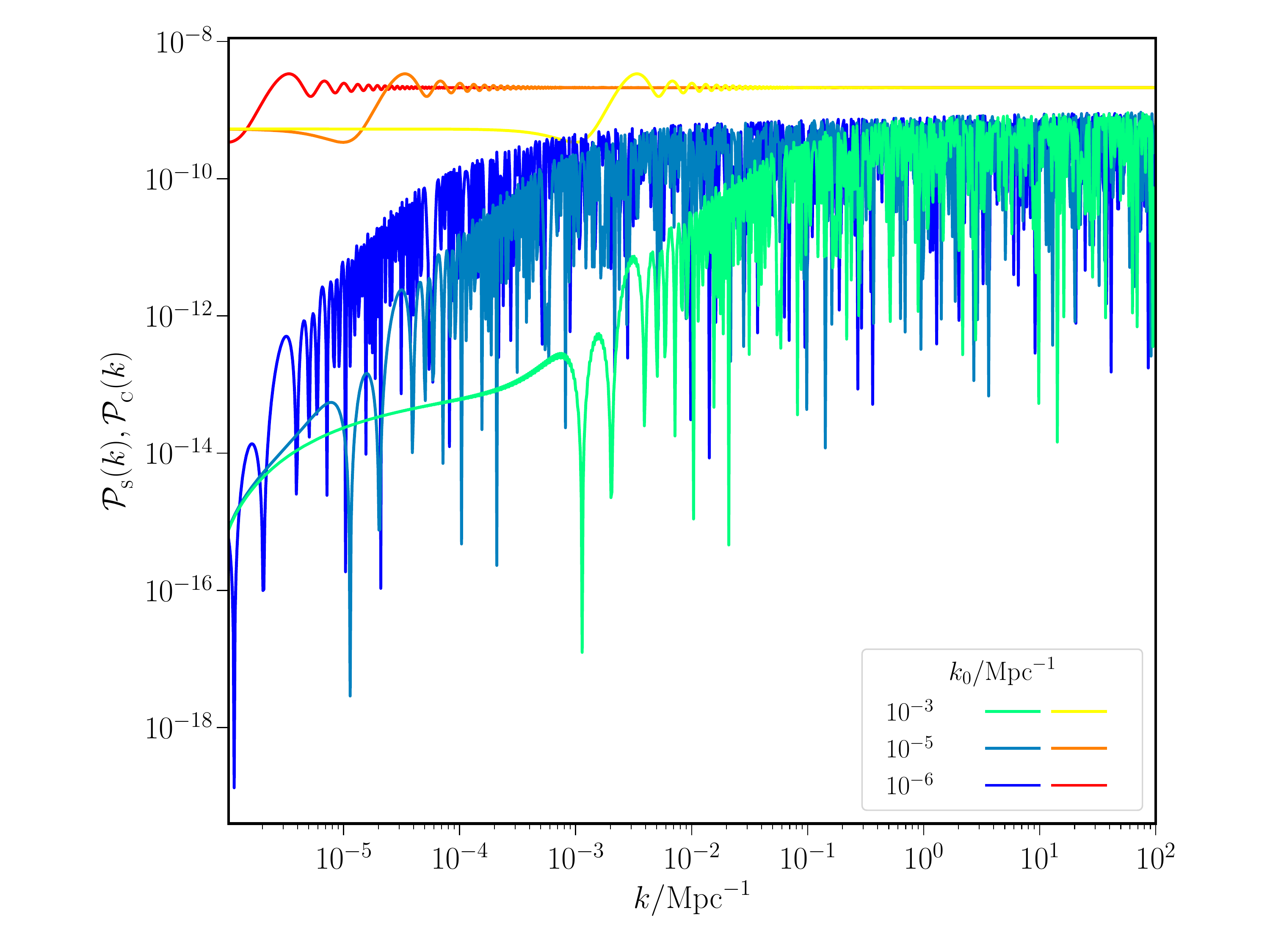}
\vskip -0.1in
\caption{We present $\ps(k)$ (in shades of red to yellow) and $\pc(k)$ (in shades of blue to green) arising from Starobinsky model for a range of the parameter $k_0/\,\mpcinv$. We have set $V_0/\Mpl^4 =2.48 \times 10^{-12}$, $(A_-/V_0)/\Mpl^{-1}=3.14 \times 10^{-3}$ and $(A_+/V_0)/\Mpl^{-1}=6.28 \times 10^{-3}$, in obtaining these plots. As $k_0$ is increased, the oscillatory patterns shift to right in both 
$\ps(k)$ and $\pc(k)$ as expected.}
\label{fig:PsPc_StaroII_k0}
\end{figure}
For discussing the results of this model, we shall consider the four parameters,
namely, $V_0/\Mpl^4$, $(A_-/V_0)\Mpl$, $\Delta A/A_+$ and $k_0/\mpcinv$. 
From Eq.~\eqref{eq:ps_staro}, it is clear that $V_0/\Mpl^4$ and 
$(A_-/V_0)\Mpl$ only affect the amplitude of $\ps(k)$ which is already constrained 
by data to be about $2.1 \times 10^{-9}$. The combination of parameters that 
dictate the prominence of features in $\ps(k)$ is $\Delta A/A_+$, whereas $k_0$ 
decides the location of these features in the range of wavenumbers. 
On the other hand, $\pc(k)$ is proportional not only to $V_0/\Mpl^4$ and 
$(A_-/V_0)\Mpl$ like $\ps(k)$, but also to $A_-/A_+$ and $k_0/k$. So, we can 
expect to see a large variation in $\pc(k)$ just by changing the behaviour and 
location of oscillations in $\ps(k)$. Thus, we shall explore variation in the 
values of $A_+$ and $k_0$ and study the effect on $\ps(k)$, $\pc(k)$ and $C_\ell$ .

\begin{figure}[t]
\centering
\includegraphics[width=12cm,height=8cm]{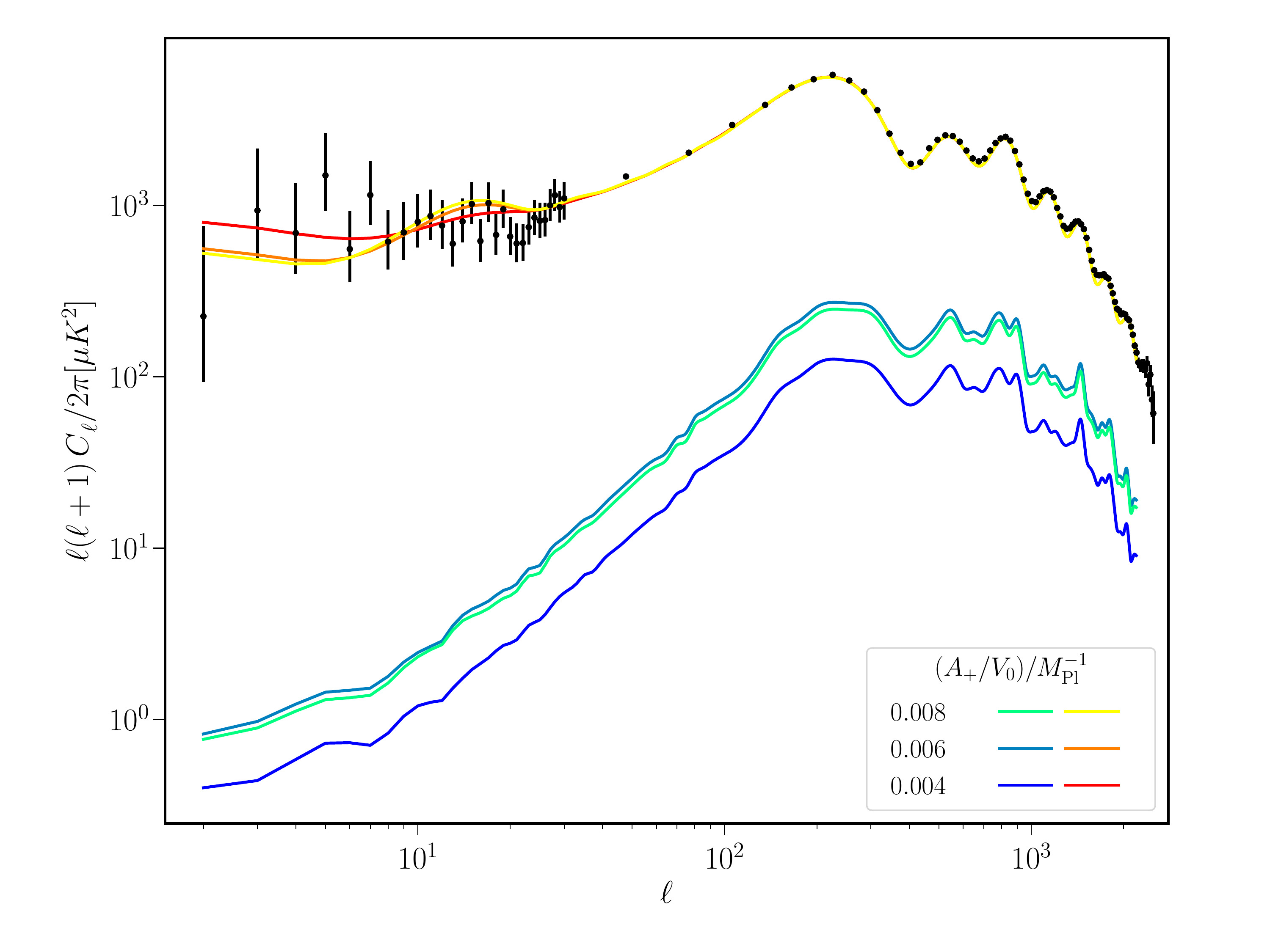}
\vskip -0.1in
\caption{The angular spectra obtained from $\ps(k)$ (in shades of red to yellow) 
and $\pc(k)$ (in blue to green) are presented for the Starobinsky model across a 
range of the parameter $(A_+/V_0)/\Mpl^{-1}$. We have set $k_0/\,\mpcinv=3.89 \times 10^{-4}$, $V_0/\Mpl^4=2.48 \times 10^{-12}$ and $(A_-/V_0)/\Mpl^{-1}=3.14 \times 10^{-3}$, in obtaining these plots. The oscillatory patterns are more pronounced in the spectra due to $\ps(k)$ as we increase this parameter. The $C_{\ell}$s due to $\pc (k)$ in general grow toward large $\ell$ values and the highest amplitude is achieved when $(A_+/V_0)/\Mpl^{-1}$ is twice that of $(A_-/V_0)/\Mpl^{-1}$.}
\label{fig:Cls_StaroII_Ap}
\end{figure}

We present the effect of variation of $(A_+/V_0)/\Mpl$ on $\ps(k)$ and $\pc(k)$ in
Fig.~\ref{fig:PsPc_StaroII_Ap}. 
We fix the values of other model parameters to be $V_0/\Mpl^4 =2.48 \times 10^{-12}$,  
$(A_-/V_0)/\Mpl^{-1}=3.14 \times 10^{-3}$ and $k_0/\,\mpcinv = 3.89 \times 10^{-4}$.
We note that the overall envelope of $\pc(k)$ is dictated by $\ln(k_{\rm min}/k)$, as 
can be expected from our examination of local type $\fnl$ where regulating the
divergence at $(x,y)=(0,1)$ lead to such a function in $\pc(k)$.
Further, the combination of terms that appear in the prefactor of $\pc(k)$, 
$(A_-/A_+)^2(1-A_-/A_+)^2$, suggests that the amplitude shall be maximum at 
$A_+ = 2\,A_-$.
We vary the value of $(A_+/V_0)\Mpl$ around this range and find that the maximum
of $\pc(k)$ is indeed achieved at $(A_+/V_0)\Mpl = 6 \times 10^{-3} 
\simeq 2(\,A_-/V_0)\Mpl$.
The corresponding CMB angular spectra are presented in Fig.~\ref{fig:Cls_StaroII_Ap}.
The standard spectra are computed for $\ps(k)$ with the corresponding values of
$(A_+/V_0)\Mpl$. We find that the amplitude of $C_\ell$s due to $\pc(k)$, though
subdominant to $C_\ell$s due to $\ps(k)$, may leave minor imprints on the latter as
their amplitudes grow over small scales of around $\ell \geq 10^3$. Moreover, a 
crucial takeaway from this effect is that, the dependence of $\pc(k)$ on $A_+$ and $A_-$,
which is distinct from the factor $\Delta A/A_+$ as it appears in $\ps(k)$, shall
help resolve the degeneracy between $A_+$ and $A_-$ if compared against the data.

\begin{figure}[t]
\centering
\includegraphics[width=12cm,height=8cm]{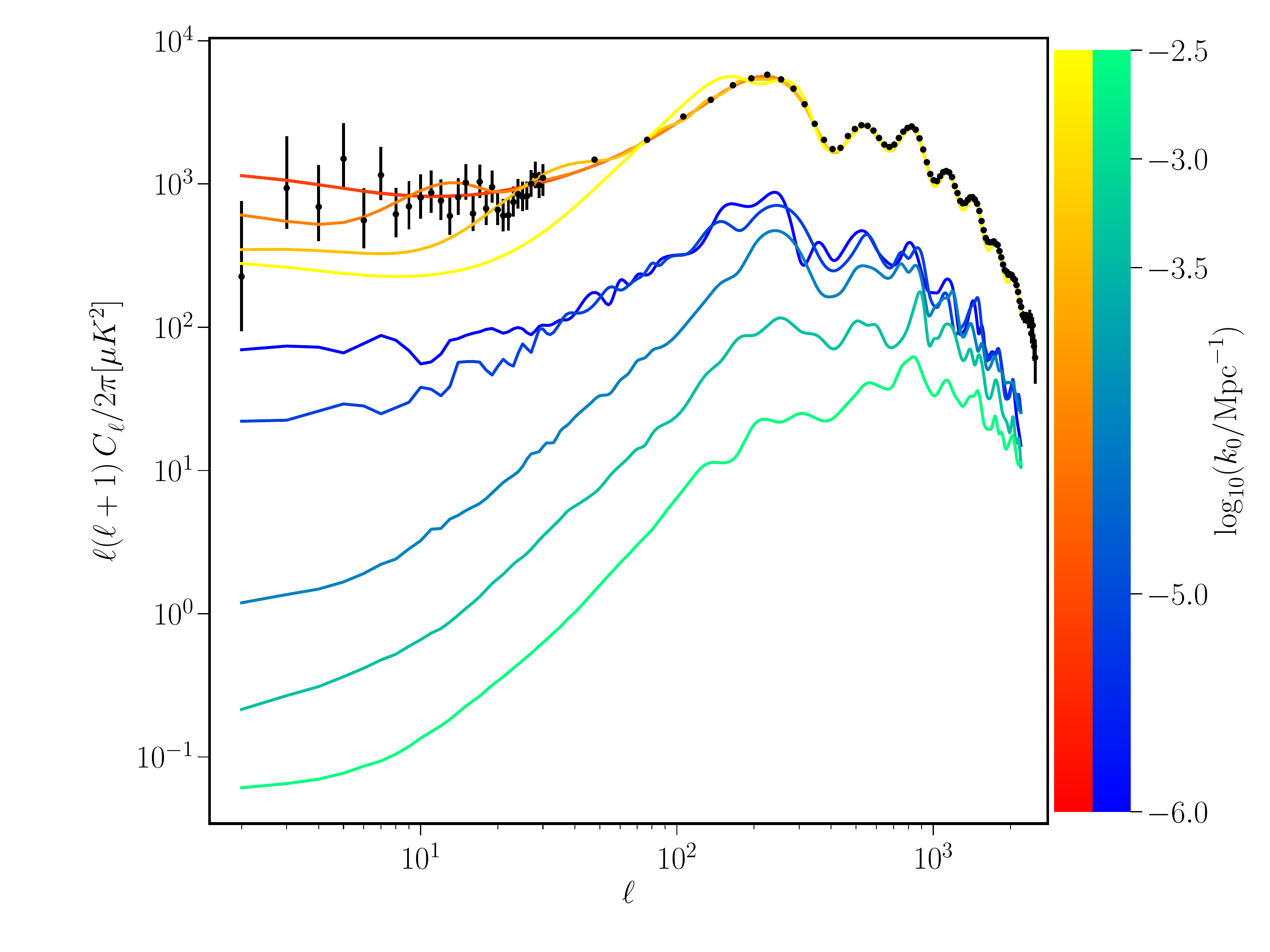}
\vskip -0.1in
\caption{The angular spectra derived from $\ps(k)$ (in red to yellow) and $\pc(k)$
(in blue to green) is presented for the Starobinsky model across a range of the parameter $k_0/\,\mpcinv$. We have set $V_0/\Mpl^4 =2.48 \times 10^{-12}$, $(A_-/V_0)/\Mpl^{-1}=3.14 \times 10^{-3}$ and $\Delta A/A_+ = -0.5 \Rightarrow (A_+/V_0)/\Mpl^{-1}=6.28 \times 10^{-3}$, in obtaining these plots. As $k_0$ is decreased, 
the oscillatory patterns shift toward small $\ell$ values in $C_\ell$s due to $\ps(k)$.
However, the amplitude of $C_\ell$s due to $\pc(k)$ increases with decrease in 
$k_0$.}
\label{fig:Cls_StaroII_k0}
\end{figure}

In Fig.~\ref{fig:PsPc_StaroII_k0}, we illustrate the effect of varying $k_0/\mpcinv$.
We fix $(A_+/V_0)/\Mpl^{-1}=6.28 \times 10^{-3}$ along with $V_0$ and $A_-$ as their values mentioned earlier.
We infer that varying $k_0$ shifts the location of the onset of features in both 
$\ps(k)$ and $\pc(k)$. The oscillations over $k > k_0$ are particularly strong
in $\pc(k)$ and they settle at asymptotic values over $k \gg k_0$ essentially, due
to the regulatory factor $1/[1 + (k+kx+ky)/k_{\rm reg}]$ introduced in the 
bispectrum.
Over $k \ll k_0$, we find that the $\pc(k)$ tend to values independent of $k_0$. 
This can be understood as an indication that the integral in 
Eq.~\eqref{eq:pc_staro2} evaluates to $(k/k_0)^2$
over $k \ll k_0$ so that the term $(k_0/k)^2$ in the prefactor is cancelled, leaving 
$\pc(k)$ independent of $k_0$.
The overall envelope of $\pc(k)$ is dictated by $\ln(k_{\rm min}/k)$ as seen before.
The corresponding angular spectra are presented in Fig.~\ref{fig:Cls_StaroII_k0}.
Surprisingly, we find that decrease in the value of $k_0$ leads to an increase in 
the amplitude of $C_\ell$s especially over large scales. However, this also 
pushes the feature of suppression in the original $C_\ell$s due to $\ps(k)$ out of
the observable window. This complementary behavior in the spectra where the amplitude
of $C_\ell$s due $\pc(k)$ is enhanced while feature in $C_\ell$s due $\ps(k)$ is
suppressed leads to an interesting case where, for $k_0/\mpcinv=10^{-6}$, the
$C_\ell$s due $\pc(k)$ acquire a magnitude of roughly about $10\%$ of the original
$C_\ell$s due $\ps(k)$.

In summary, in this model of Starobinsky, $C_{\ell}$s due to $\pc(k)$ do not become 
comparable to the standard $C_{\ell}$s, for the range of values explored in $k_0$
and $A_+$. However, we see that the former may induce $1-10\%$ change in the latter
for certain values of these parameters and lead to interesting effects, mainly because
of the difference in the dependence of $\pc(k)$ and $\ps(k)$ on the model parameters. 
This is a compelling case to consider the contribution to $C_\ell$s due to $\pc(k)$
in comparing this model against the data where the current constraints on
$k_0$ may be altered and the degeneracy between $A_-$ and $A_+$ may be 
ameliorated due to the unique dependence of $\pc(k)$ on them.

\section{Conclusion}\label{sec:conc}

This work examines a method of accounting non-Gaussian corrections to the
primordial scalar power, arising at the level of scalar bispectrum through the 
associated non-Gaussianity parameter $\fnl(k_1,k_2,k_3)$. 
We have computed the CMB angular spectra arising from such non-Gaussian 
corrections for various templates of $\fnl$ and for Starobinsky model. We have 
studied their dependence on various model parameters and determined the typical 
range of parameters that may lead to imprints on the standard angular spectrum. 
We have found that such an exercise can possibly alter the existing fitness 
of these models against the Planck dataset.

We have shown that this method provides an avenue to explore novel dependences 
on model parameters that are not captured in the original power spectrum of the 
Gaussian perturbations. This property is well observed in the cases of oscillatory 
template of power and bi-spectra and Starobinsky model. They can, in principle, 
break, or at least reduce, the degeneracies amongst parameters, which cannot 
be resolved at the level of $\ps(k)$.

However, there are a few caveats and hence room for improvement in this analysis.
Since we have focused on illustrating the major effects of $\pc(k)$ on CMB, we have 
not completely computed the complicated integrals over ($x,y)$, that arise in 
various templates of $\fnl(k_1,k_2,k_3)$. Since, they are of ${\cal O}(1)$, we 
have either worked with their maximum possible contribution or set them to unity 
to arrive at our results.
As a future course of this work, we intend to carry out these integrals exactly, 
either analytically or numerically. Particularly, we shall focus on realistic 
models of inflation along the lines of Starobinsky model, and compute the 
complete $\pc(k)$ arising from them. We shall account for the non-linear lensing,
that has been neglected in this work, while studying the complete angular 
spectra due to $\ps(k)+\pc(k)$.
More importantly, we plan to perform a proper Bayesian analysis of comparing the 
complete angular spectra arising from these models, against the latest dataset 
of Planck 2018, comprising of both temperature and polarization spectra. We 
intend to study the updated posteriors of the associated parameters and infer 
any improvement or worsening in the fit of certain promising models due to their 
non-Gaussianities~\cite{Hazra:2014goa,Ragavendra:2020old,Antony:2022ert}. 
We are presently working in this direction.

Furthermore, this method can be extended to compute non-Gaussian corrections to
the scalar power due to other types of three-point correlations such as 
scalar-tensor-tensor and scalar-scalar-tensor 
types~\cite{Sreenath:2013xra,Chowdhury:2016yrh}.
It can be used to examine and constrain models with spectator fields that strongly 
interact with the scalar perturbations and thus giving rise to significant 
levels of non-Gaussianities~\cite{Wang:2021qez,Chen:2022vzh}.
It can also be used to capture the effect of cross-correlation between 
scalar perturbations and gauge fields, in models of inflation that are considered
in the context of primordial magnetogenesis~\cite{Chowdhury:2018mhj,Tripathy:2021sfb,
Tripathy:2022iev}.
In summary, the method presented in our work shall serve as an effective tool 
for examining and constraining a variety of non-Gaussianities arising in some of
the non-trivial models of inflation.


\acknowledgments
BD thanks Dr.~Koushik Dutta for useful comments and suggestions. 
HVR thanks Raman Research Institute for support through postdoctoral research 
fellowship.
BD and HVR thank International Centre for Theoretical Sciences (ICTS) for 
hospitality during the program titled Less Travelled Path to the Dark Universe 
(code:~ICTS/ltpdu2023/3), where a part of this work was completed.

\appendix


\section{Behavior of integrands}\label{app:integrands}

The behaviors of integrands involved in arriving at $\pc(k)$ arising in various 
templates and the model studied are presented and briefly discussed in this 
appendix. We plot the the density map of these integrand over a range of the 
variables of integration $x$ and $y$. Since the integrands typically decay 
rapidly as $1/(x^2y^2)$, we focus on region around origin to study their shapes.


\subsection{Local type}

In case of local type $\fnl$, the integrand diverges at the two points of
$(x,y)=(0,1)$ and $(1,0)$ due to the presence of $1/(xy)^2$ term. The computation
of the double integrals using a finite value of $k_{\rm min}$ can be understood
as follows. Using Eq.~\eqref{eq:pc-fnl}, we may obtain
\begin{eqnarray}
\pc(k) &=& \f{9}{25}(\fnl^{\rm loc})^2\,
A_s^2\,\left(\f{k}{k_\ast} \right)^{2(\ns-1)}
\int_0^\infty {\rm d}x\, x^{\ns-3}
\int_{\vert 1 - x \vert}^{(1+x)} {\rm d}y\, y^{\ns-3}.
\end{eqnarray}
We shall ignore the minor effect of $x^{\ns-1}$ and $y^{\ns-1}$ within the 
integral and rewrite $\pc(k)$ as
\begin{eqnarray}
\pc(k) &=& \f{9}{25}(\fnl^{\rm loc})^2 \ps^2(k) \int_0^\infty {\rm d}x\, x^{-2} \int_{\vert 1 - x \vert}^{(1+x)} {\rm d}y\, y^{-2} \nn \\ 
&=& -\f{9}{25}(\fnl^{\rm loc})^2 \ps^2(k) \int_0^\infty {\rm d}x\, x^{-2} \left(\f{1}{1+x}-\f{1}{\vert 1 - x \vert}\right) \nn \\
&=& -\f{9}{25}(\fnl^{\rm loc})^2 \ps^2(k) \left\{\int_0^1 {\rm d}x\, x^{-2} \left(\f{1}{1+x}-\f{1}{1-x}\right) + \int_1^\infty {\rm d}x\, x^{-2} \left(\f{1}{1+x}+\f{1}{1-x}\right)\right\} \nn \\
&=& \f{18}{25}(\fnl^{\rm loc})^2 \ps^2(k) \left\{\int_0^1 {\rm d}x\, \f{1}{x(1-x^2)} + \int_1^\infty {\rm d}x\, \f{1}{x^2(x^2-1)}\right\} \nn \\
&=& \f{18}{25}(\fnl^{\rm loc})^2 \ps^2(k) \lim_{x_{\rm min} \rightarrow 0} \left\{\ln{\vert 1 + x_{\rm min}\vert} - \f{\ln{\vert (1 + x_{\rm min})^2 -1 \vert}}{2} - \ln{\vert x_{\rm min}\vert} \right. \nn \\
& & + \left. \f{\ln{\vert x_{\rm min}^2 -1 \vert}}{2} + \f{\ln{(2+x_{\rm min})}}{2} - \f{1}{1+x_{\rm min}} - \f{\ln{x_{\rm min}}}{2}\right\} \nn \\
&=& \f{18}{25}(\fnl^{\rm loc})^2 \ps^2(k) \lim_{x_{\rm min} \rightarrow 0} \left\{\f{3\ln{(1+x_{\rm min})}}{2} - 2\ln{x_{\rm min}} + \f{\ln{(1-x_{\rm min})}}{2} - \f{1}{1+x_{\rm min}}\right\} \nn \\
&=& -\f{18}{25}(\fnl^{\rm loc})^2 \ps^2(k) \left(1 + 2\lim_{x_{\rm min} \rightarrow 0} \ln{x_{\rm min}} \right).
\end{eqnarray}
The quantity $x_{\rm min} = {k_{\rm min}}/{k}$ and so
\begin{eqnarray}
\pc(k) &=& -\f{18}{25}(\fnl^{\rm loc})^2 \ps^2(k) \left(1 + 2\lim_{k_{\rm min} \rightarrow 0} \ln{\f{k_{\rm min}}{k}} \right)\,.
\end{eqnarray}
Here and wherever such regulation is required, the value of $k_{\rm min}/\,\mpcinv$ is 
set to be $10^{-6}$ as mentioned in the main text. The shape of 
$\pc(k)$ is hence detemined by $(k/k_\ast)^{2(\ns-1)}$ in $\ps(k)$ and
$\ln(k_{\rm min}/k)$.


\subsection{Orthogonal type}

\begin{figure}[t]
\centering
\includegraphics[width=12cm,height=8cm]{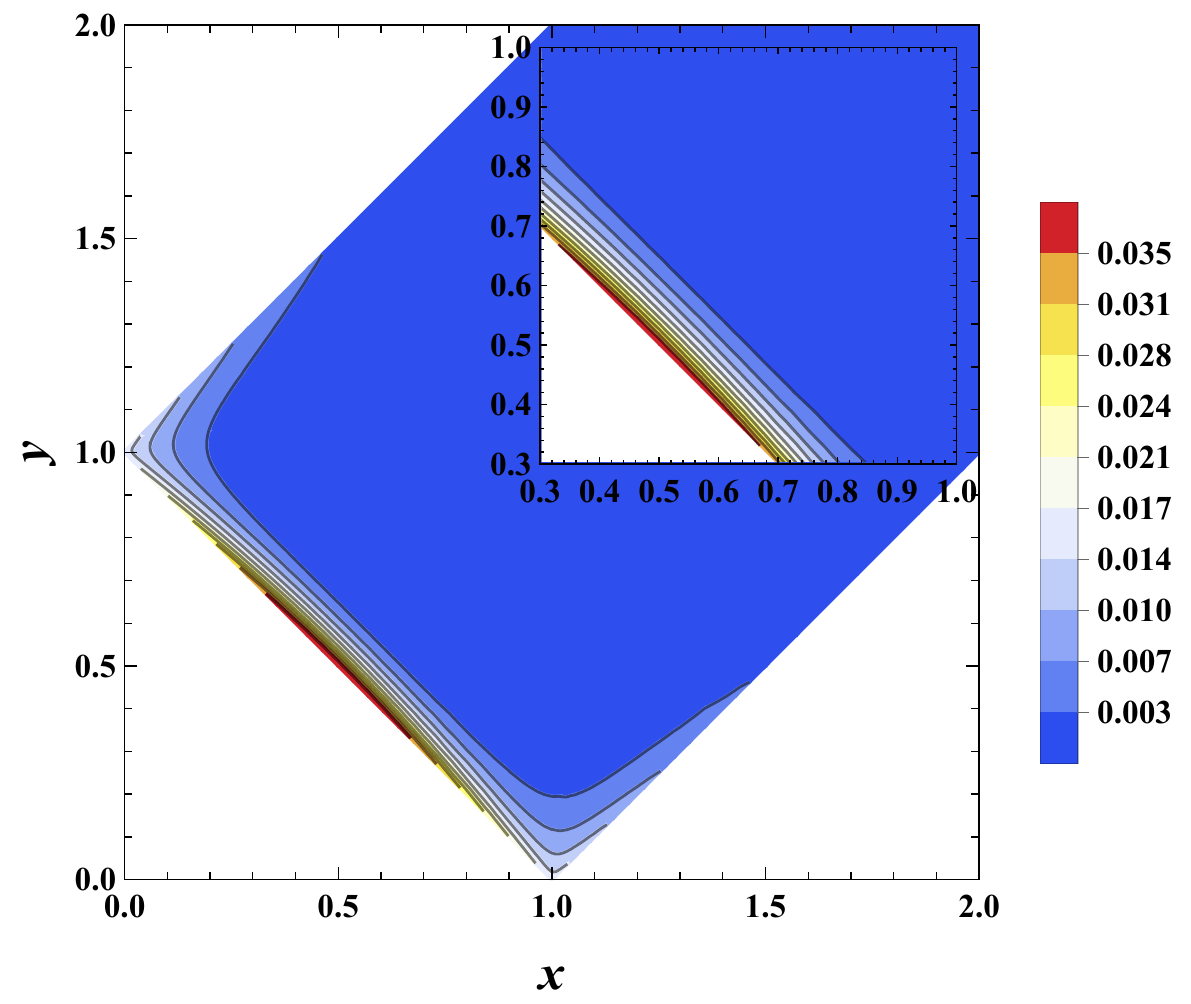}
\vskip -0.1in
\caption{The integrand involved in $\pc^{\rm ortho}(k)$ is plotted as a function of $x,y$. We have separated and accounted for the constants namely, $A_{_{\rm S}}$ and $\fnl^{\rm ortho}$ in $\pc^{\rm ortho}(k)$ and plot here, only the part of function dependent on $x,y$. We find that the integrand peaks along $y=1-x$, with maximum at $x=y=1/2$ and falls rapidly for larger values of $x$ and $y$.}
\label{fig:Integrand_ortho}
\end{figure}

In case of orthogonal form of $\fnl(k_1,k_2,k_3)$, the integrand that arises in 
calculating $\pc(k)$ is sharply peaked along the line of $y=1-x$. This is 
depicted in Fig.~\ref{fig:Integrand_ortho}. But the integrand does not diverge 
as it happens in the case of local type. So we focus in integrating over this 
region to capture the dominant contribution to $\pc(k)$.


\subsection{Equilateral type}

\begin{figure}[h]
\centering
\includegraphics[width=12cm,height=8cm]{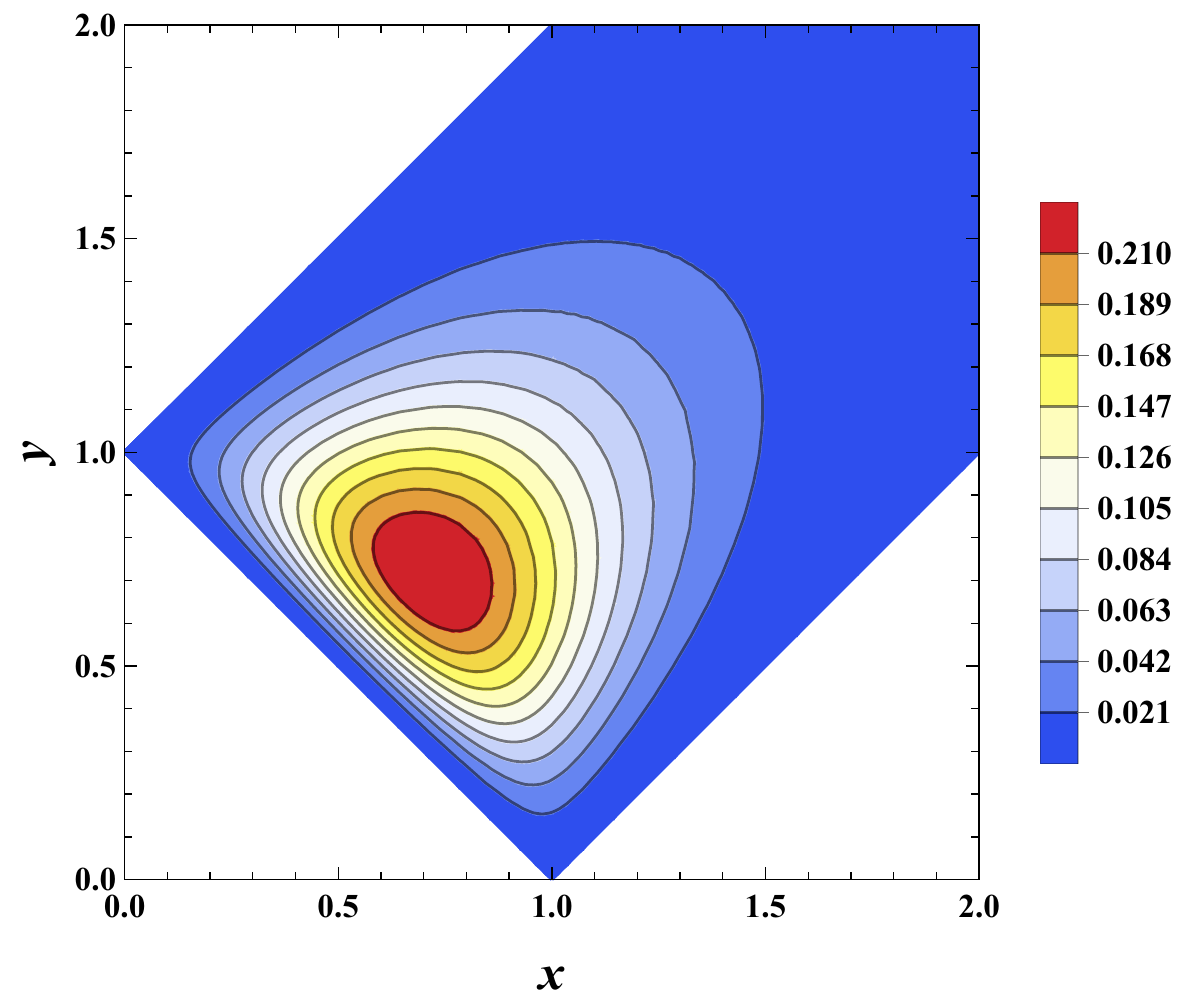}
\vskip -0.1in
\caption{The integrand involved in $\pc^{\rm eq}(k)$ is plotted as a function of $x,y$. As in the previous plot, we have already accounted for the constants like $A_{_{\rm S}}$ and $\fnl^{\rm eq}$. We find that the integrand has a maximum at $x=y=0.719$ and falls rapidly away from this region.}
\label{fig:Integrand_eq}
\end{figure}

In case of equilateral shape of $\fnl(k_1,k_2,k_3)$, the integrand once again
is well behaved without any divergences as plotted in Fig.~\ref{fig:Integrand_eq}.
The amplitude is localized and of order unity around the region of $x=y=1$, and 
it rapidly decays down to smaller values away from this region. Hence, we
perform integration over this region to obtain the magnitude of the 
contribution to $\pc(k)$.


\subsection{Oscillatory type}

\begin{figure}[t]
\centering
\includegraphics[width=12cm,height=8cm]{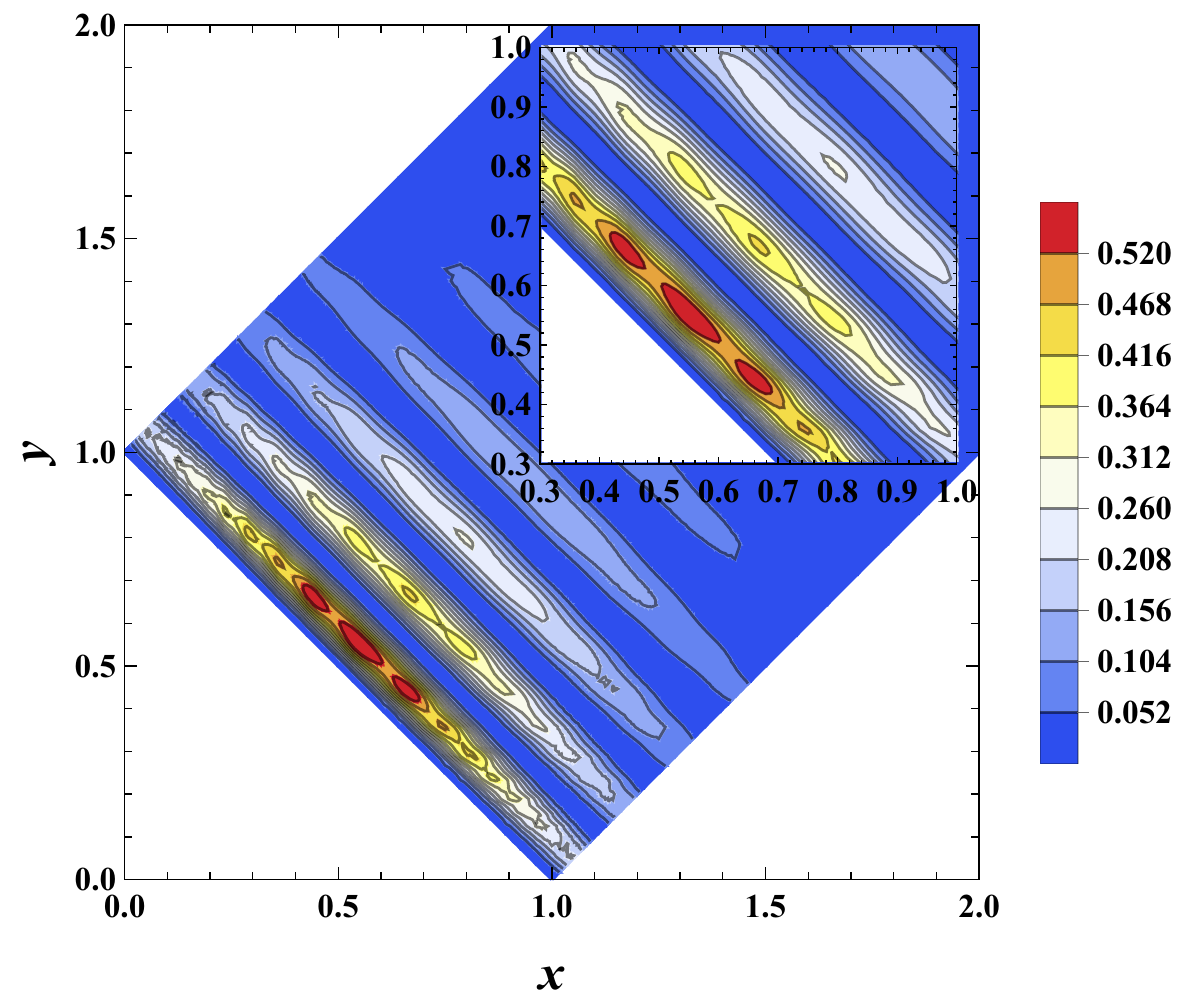}
\vskip -0.1in
\caption{The integrand involved in computing $\pc^{\rm osc}(k)$ is plotted as a 
function of $x,y$. Note that we have plotted only the integrand involving $x,y$ 
excluding the numerical factors such as $A_{_{\rm S}}$, $\fnl^{\rm osc}$ 
[cf.~Eq.~\eqref{eq:pc-osc-full}].
We have chosen other parameters to be $k=k_\ast,\,k_o/\,\mpcinv=0.1$ and $b=0.05$.
We have set $\omega=30$ to better illustrate the oscillatory patterns in $x$ as
well as $y$ directions in the range presented. As can be expected, the quantity 
is positive throughout the range and unity at maximum. It peaks around $x=y=0.55$ 
and falls rapidly over large values of $x,y$ as can be seen in the inset.}
\label{fig:Integrand_osc}
\end{figure}

We present the integrand for the oscillatory template of $\fnl(k_1,k_2,k_3)$ in 
Fig.~\ref{fig:Integrand_osc}. We see that the integrand exhibits oscillatory 
patterns along both $x$ and $y$ directions. There are no divergences present and
further, it decays down in amplitude as $x$ and $y$ grow to large values. 
Note that the integrand plotted corresponds to the complete expression as 
described in Eq.~\eqref{eq:pc-osc-full}.
However, we have expanded this integral order by order in the parameter $b$ and 
extracted the scale dependence at each order in the main text.
The associated integrals, denoted as $\cal I$ and ${\cal I}^{(b)}$, that arise 
at the level of ${\cal O}(b^0)$ and ${\cal O}(b)$ respectively, are presented 
below.
\begin{figure}[t]
\centering
\includegraphics[width=12cm,height=8cm]{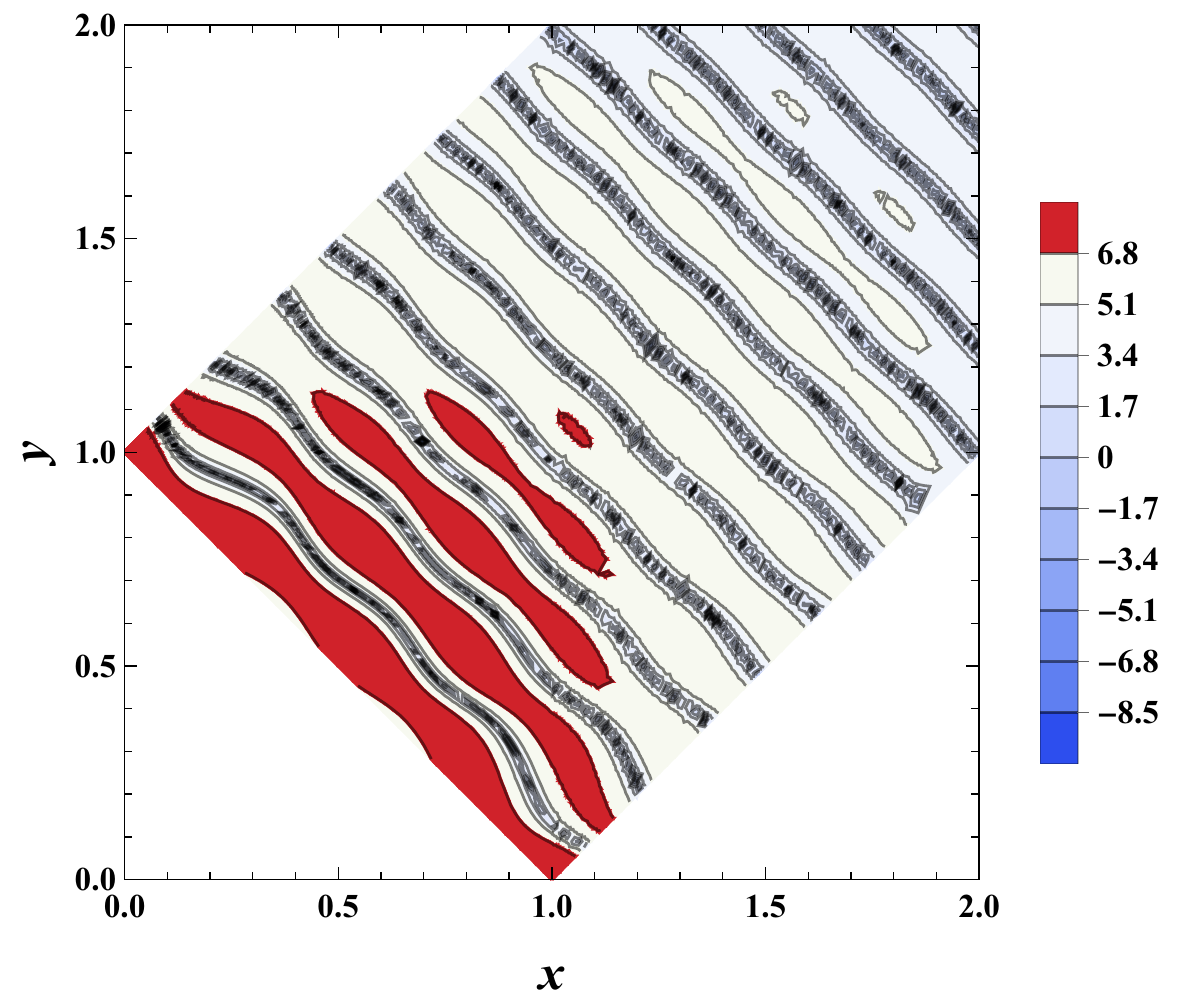}
\includegraphics[width=9cm,height=6cm]{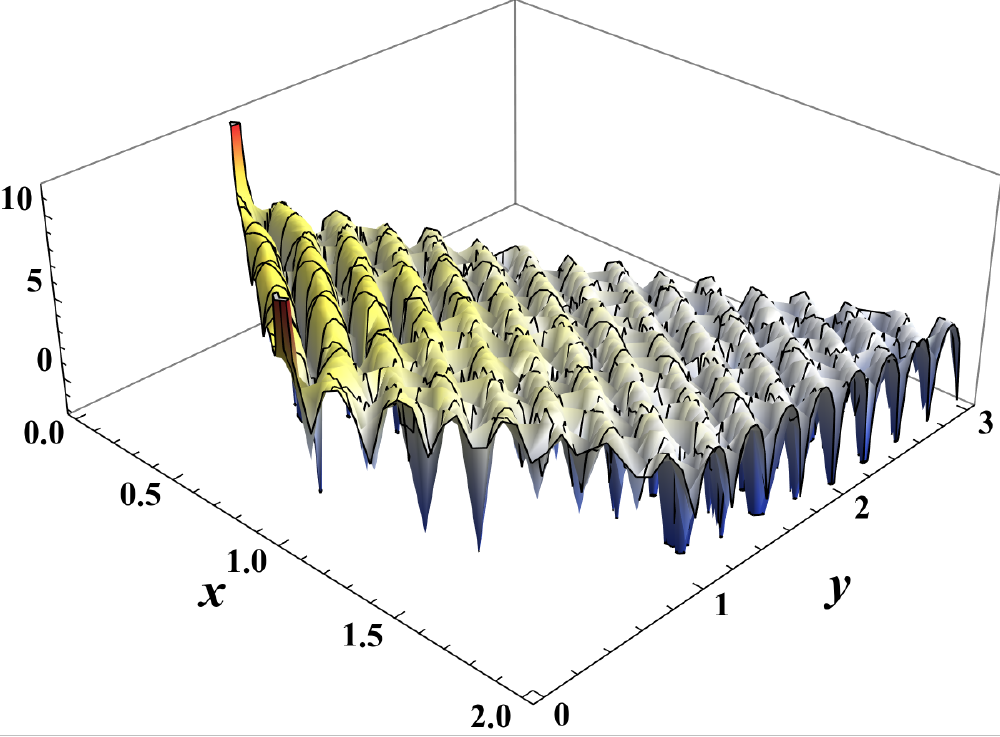}
\vskip -0.1in
\caption{The integrand involved in computing $\pc(k)$ for Starobinsky model is 
plotted as a function of $x,y$, as a 2D projection on top and 3D plot below. We 
have plotted only the integrand involving $x,y$ excluding the numerical scaling 
factor that is $(V_0/\Mpl^4)^2/[(A_-/V_0)/\Mpl^{-1}]^4$ arising due to 
$\ps(kx),\,\ps(ky)$. We have chosen parameters to be $k=k_\ast, k_o/\,\mpcinv=4 \times 10^{-3}$,$V_0/\Mpl^4 =2.48 \times 10^{-12}$ and $(A_-/V_0)/\Mpl^{-1}=3.14 \times 10^{-3}$ and $\Delta A/A_+ = -0.5$.
We have plotted the integrand in natural log to better illustrate the complex 
oscillations in $x$ as well as $y$ directions in the range presented. We find 
that it diverges at $(x,y)=\{(0,1),(1,0)\}$ and falls rapidly over large 
values of $x,y$ as seen in the 3D plot below.}
\label{fig:Integrand_staro}
\end{figure}

The integrals involved in the expression of $\pc^{\rm osc}(k)$ at the level of
${\cal O}(b^0)$ are [cf.~Eq.~\eqref{eq:pc-b0}]
\begin{eqnarray}
{\cal I}_1 &=& \int_0^\infty{\rm d}x\int^{1+x}_{\vert 1-x\vert}{\rm d}y
\,\frac{1}{(1+x^3+y^3)^2}\,, \\
{\cal I}_2 &=& \int_0^\infty{\rm d}x\int^{1+x}_{\vert 1-x\vert}{\rm d}y
\,\frac{\cos [2\omega\ln(1+x+y)] }{(1+x^3+y^3)^2}\,, \\
{\cal I}_3 &=& \int_0^\infty{\rm d}x\int^{1+x}_{\vert 1-x\vert}{\rm d}y
\,\frac{\sin [2\omega\ln(1+x+y)] }{(1+x^3+y^3)^2}\,.
\end{eqnarray}

The integrals required in computation of $\pc^{\rm osc}(k)$ at the level of
${\cal O}(b)$ are [cf.~Eq.~\eqref{eq:pc-b1}]
\begin{eqnarray}
{\cal I}^{(b)}_1 &=& \int^\infty_0 {\rm d}x \int^{1+x}_{\vert 1-x\vert} {\rm d}y
\cos^2\left[\omega \ln(1+x+y)\right]\,\frac{f_1(x,y)}{(1+x^3+y^3)^3}\,, \\
{\cal I}^{(b)}_2 &=& \int^\infty_0 {\rm d}x \int^{1+x}_{\vert 1-x\vert} {\rm d}y
\cos^2\left[\omega \ln(1+x+y)\right]\,\frac{f_2(x,y)}{(1+x^3+y^3)^3}\,, \\
{\cal I}^{(b)}_3 &=& \int^\infty_0 {\rm d}x \int^{1+x}_{\vert 1-x\vert} {\rm d}y
\sin^2\left[\omega \ln(1+x+y)\right]\,\frac{f_1(x,y)}{(1+x^3+y^3)^3}\,, \\
{\cal I}^{(b)}_4 &=& \int^\infty_0 {\rm d}x \int^{1+x}_{\vert 1-x\vert} {\rm d}y
\sin^2\left[\omega \ln(1+x+y)\right]\,\frac{f_2(x,y)}{(1+x^3+y^3)^3}\,, \\
{\cal I}^{(b)}_5 &=& \int^\infty_0 {\rm d}x \int^{1+x}_{\vert 1-x\vert} {\rm d}y
\sin2\left[\omega \ln(1+x+y)\right]\,\frac{f_1(x,y)}{(1+x^3+y^3)^3}\,, \\
{\cal I}^{(b)}_6 &=& \int^\infty_0 {\rm d}x \int^{1+x}_{\vert 1-x\vert} {\rm d}y
\sin2\left[\omega \ln(1+x+y)\right]\,\frac{f_2(x,y)}{(1+x^3+y^3)^3}\,,
\end{eqnarray}
where the functions $f_1(x,y)$ and $f_2(x,y)$ are given by
\begin{eqnarray}
f_1(x,y) &=& x^3 + y^3 + (1+x^3)\cos[\omega\ln y] + (1+y^3)\cos[\omega\ln x]\,, \\
f_2(x,y) &=& (1+x^3)\sin[\omega\ln y] + (1+y^3)\sin[\omega\ln x]\,.
\end{eqnarray}
The integrands of these integrals and hence the integrals themselves evaluate
to ${\cal O}(1)$ and hence are assumed to be unity in illustration of 
$\pc^{\rm osc}(k)$ and the associated $C_\ell$s.
\subsection{Starobinsky model}
The integrand involved in calculation of $\pc(k)$ for Starobinsky model is
presented in Fig.~\ref{fig:Integrand_staro}. We see that it has a highly
non-trivial oscillatory feature throughout the range of integration. There
are divergences at $(x,y)=(0,1)$ and $(1,0)$ as was the case for local type.
Moreover, there is a decay of amplitude as $x$ and $y$ tend toward large values.
We have also presented the same integrand in a 3D plot, projected at an angle,
to better illustrate its nature. The complicated shape of the integrand does
not allow for any approximate evaluation of the integral and hence we perform
the integration numerically.


\bibliographystyle{apsrev4-2}
\bibliography{ng_ps_cmb}

\end{document}